\documentclass{article}

\usepackage[nonatbib, final]{auxiliary/neurips_2023}

\usepackage[square,numbers]{natbib}
\usepackage{auxiliary/arxivpackages}
\usepackage{auxiliary/mymacros}
\usepackage{auxiliary/commands}
\usepackage{auxiliary/filecommands}

\newcommand{\mytitle}[1][]{Convex-Concave Zero-Sum \sdeni{Stochastic}{Markov} Stackelberg Games}
\newcommand{\mysubtitle}[1][]{}
\newcommand{\myauthors}[1][]{Denizalp Goktas}
\newcommand{\myauthorsshort}[1][]{Goktas}

\usepackage[utf8]{inputenc} 
\usepackage[T1]{fontenc}    
\usepackage{hyperref}       
\usepackage{url}            
\usepackage{booktabs}       
\usepackage{amsfonts}       
\usepackage{nicefrac}       
\usepackage{microtype}      
\usepackage{xcolor}         

\title{\mytitle}

\author{%
  Denizalp Goktas\\
  Brown University, Computer Science\\
  \texttt{denizalp\_goktas@brown.edu} \\
  \And
   Arjun Prakash\\
  Brown University, Computer Science\\
  \texttt{arjun\_prakash@brown.edu} \\
   \And
  Amy Greenwald\\
  Brown University, Computer Science\\
   \texttt{amy\_greenwald@brown.edu} \\
}

\begin{document}

\maketitle

\begin{abstract}
Zero-sum Markov Stackelberg games can be used to model myriad problems, in domains ranging from economics to human robot interaction.
In this paper, we develop policy gradient methods that solve these games \sdeni{}{in continuous state and action settings} using noisy gradient estimates computed from observed trajectories of play.
When the games are convex-concave, we prove that our algorithms converge to 
Stackelberg equilibrium in polynomial time.
We also show that
reach-avoid problems are naturally modeled as convex-concave zero-sum Markov Stackelberg games, and 
that Stackelberg equilibrium policies are more effective
than their Nash counterparts in these problems.

\if
    We propose novel policy gradient methods to solve convex-concave 
    zero-sum stochastic Stackelberg games, and we prove that our methods converge to recursive (i.e., Markov perfect) Stackelberg equilibrium in polynomial time when the games are convex-concave.
    We showcase our method by applying it to solve reach-avoid problems, which naturally satisfy the requisite convex-concave condition. 
    We also present simulation experiments that demonstrate the effectiveness of our method.
    Specifically, we show that viewing these problems as Stackelberg (i.e., sequential) rather than Nash (i.e., simultaneous) games improves safety without sacrificing liveness.
\fi

\end{abstract}

\deni{TO FIX:
\begin{enumerate}
    \item the distinction between $\numhorizon$ and $\numiters$
    \item $\constrdistrib$ and $\objdistrib$ notation inconsistency with conventions.
    \item Fix proof of lemma 2.
    \item reward and payoff 
\amy{need to do a full search and replace for rewards to payoffs}
    \item change coupled policy correspondence to $\coupledset$.
    \item fix stochastic convexity assumption for lemma 1 and 2 on trans prob.
    \item softmax parameterization 
    \item add log-convex-log-concave game characterization result.
    \item Use acronyms GDA and SGDA everywhere. 
    \item Fix Algos.
    \item There should be no reference to stochastic min-max optimization, only min-max optimization. \amy{i don't understand this. MDPs are stochastic optimization problems.}
    \item Future work should explore convergence guarantees for invex-incave games... etc..
    \item Make sure dimensionality of parameter space is $\numparams$.
    \item Make sure "follower's action correspondence" is defined with that name and/or standardize language.
    \item Make sure definition of Stackelberg eqm is approximately feasible.
    \begin{itemize}
        \item add objective oracle notation, and change gradient oracle notation
        \item change SE notation so that delta include constraint violation as well
        \item constraint violation bound follows directly from saddle-point oracle value.
    \end{itemize}
    \item $\coupledset$ being convex-valued is not enough for convergence $\constr(\outer, \inner)$ has to be concave in $\inner$.
    \item Change target state macro
    \item Definition of saddle point
    \item Change Lagrange multipliers to KKT multipliers.
    \item Convexity assum on $\innerset$
    \item Gendered language for the leader and follower (her/him), just convert all to it.
\end{enumerate}
}
\section{Introduction}  
 
Markov games \cite{fink1964equilibrium,shapley1953stochastic,takahashi1964equilibrium} are 
a generalization of Markov decision processes (MDPs) comprising multiple players simultaneously making 
decisions over time, collecting rewards along the way depending on their collective actions.
They have been used by practitioners to model many real-world multiagent planning and learning environments, such as autonomous driving \cite{fridovich2020efficient, palanisamy2020multi}, 
cloud computing \cite{wang2019multi}, and telecomunications \cite{altman1994flow}.
Moreover, 
theoreticians are beginning to formally analyze policy gradient methods, proving polynomial-time convergence to optimal policies in MDPs \cite{agarwal2021theory, bhandari2019global},
and to Nash equilibrium policies \cite{nash1950existence} in zero-sum Markov games \cite{daskalakis2020independent}, the canonical solution concept.
While Markov games are a fruitful way to model some problems (e.g., robotic soccer \cite{littman1994markov}),
other problems, such as reach-avoid \cite{margellos2011hamilton}, may be more productively modeled as sequential-move games, in which some players may commit to moves that are observed by others, before they make their own moves. 
To this end, we study two-player zero-sum Markov Stackelberg \cite{stackelberg1934marktform} (i.e., sequential-move) games.
\amy{what do we want to say was done in the 2022 NeurIPS paper, and what was not done? VI vs. pol grad? or exact oracle access? or what?}
Whereas \sdeni{}{polynomial-time} value-iteration (i.e., dynamic programming) algorithms are known for these games \cite{goktas2022zero} \sdeni{}{when the state space is discrete and certain convexity assumptions are satisfied}\amy{do we want to add something about the complexity of the VI algo(s)?}, we develop \sdeni{}{polynomialt-time} policy gradient methods that converge to Stackelberg equilibrium in \sdeni{these}{continuous state and action} games,\amy{do we want to add something about the complexity of our algo(s)?} using noisy gradients based only on observed trajectories of play.
Furthermore, we implement our approach to demonstrate experimentally that Stackelberg equilibrium policies are more effective than their Nash counterparts in reach-avoid problems.

A \mydef{(discounted discrete-time) zero-sum Markov}
\mydef{Stackelberg game} \cite{goktas2022zero} is
played over an infinite horizon $\numhorizon = 0, 1, \hdots$ between two players, a leader and a follower.
The game starts at time $\numhorizon = 0$, at some initial state $\staterv[0] \sim \initstates \in \simplex (\states)$ drawn randomly from a set of states $\states$.
At each subsequent time step $\iter = 1, 2, \hdots$, the players encounter a new state $\state[\iter] \in \states$, where the leader takes its action $\outeraction[][][][\iter]$ first, from its action space  
$\outeractions (\state[\iter]) \subseteq \R^\numactions$,
after which the follower, having observed the leader's action, makes it own move $\inneraction[][][][\iter]$ from a space of actions $\coupledactions (\state[\iter], \outeraction[][][][\iter]) \subseteq \inneractions(\state[\iter]) \subseteq \R^\numactions$ determined by the leader's action $\outeraction[][][][\iter]$.%
\footnote{To simplify notation, we drop the dependency of action spaces on states going forward, but our theory applies in this more general setting.} 
After both players have taken their actions, they receive respective \sdeni{payoffs}{reward}, $- \reward (\state[\iter], \outeraction[][][][\iter], \inneraction[][][][\iter])$ and $\reward (\state[\iter], \outeraction[][][][\iter], \inneraction[][][][\iter])$.
The game then transitions, either to a new state $\staterv[\iter+1] \sim \trans (\cdot \mid \state[\iter], \outeraction[][][][\iter], \inneraction[][][][\iter])$ with probability $\discount$, called the \mydef{discount factor}, or the game ends with the remaining probability.
Each player's goal is to play a (potentially non-stationary) \mydef{policy}, i.e., a sequence of history-dependent actions $\{\outeraction[][][][\iter]\}_{\iter = 0}^\infty$ (resp.\ $\{\inneraction[][][][\iter]\}_{\iter = 0}^\infty$),
that maximizes her respective \mydef{expected cumulative discounted payoffs}, $- \mathop{\Ex} \left[\sum_{\iter = 0}^\infty \discount^\iter \reward(\staterv[\iter], \outeractionrv[\iter], \inneractionrv[\iter]) \right]$ and $\mathop{\Ex} \left[\sum_{\iter = 0}^\infty \discount^\iter \reward(\staterv[\iter], \outeractionrv[\iter], \inneractionrv[\iter]) \right]$.%
\footnote{Unlike $\outeraction[][][][\iter]$ and $\inneraction[][][][\iter]$, which are deterministic actions because they depend on a realized history of states and actions encountered, $\outeractionrv[\iter]$ and $\inneractionrv[\iter]$ are random variables, because they might depend on a random history.}

In zero-sum Markov Stackelberg games, when for all $\state \in \states$, $\reward (\state, \outeraction, \inneraction)$ is continuous in $(\outeraction, \inneraction)$ and bounded; and
$\coupledactions (\state, \outeraction)$ is continuous in $\outeraction$, as well as non-empty- and compact-valued; a \mydef{recursive} (or \mydef{Markov perfect}) \cite{maskin2001markov} \mydef{Stackelberg equilibrium} is guaranteed to exist \cite{goktas2022zero}, meaning a \mydef{stationary policy profile} (i.e., a pair of mappings from states to the actions of the leader and the follower, respectively) specifying the actions taken at each state s.t.\ the leader's policy maximizes its expected payoff assuming the follower best responds, while the follower indeed best responds to the leader's policy.
%
%
In other words, under the aforementioned assumptions, we are guaranteed the existence of a \mydef{policy profile} $\policy[][][*] \doteq (\policy[\outeraction][][*], \policy[\inneraction][][*])$, with $\policy[\outeraction][][*]: \states \to \outeractions$ and $\policy[\inneraction][][*]: \states \to \inneractions$, that solves the following optimization problem:%
\footnote{We note that solutions to this problem cannot be interpreted as generalized Nash equilibria (see \cite{goktas2021minmax}).}
\begin{align}
\label{eq:min_max_stoch_stackelberg}
\min_{\policy[\outeraction]: \states \to \outeractions} \max_{\substack{\policy[\inneraction]: \states \to \inneractions:\\ \forall \state \in \states, \policy[\inneraction](\state) \in \coupledactions(\state, \policy[\outeraction](\state))}} \Ex \left[\sum_{\iter = 0}^\infty \discount^\iter \reward(\staterv[\iter], \policy[\outeraction](\staterv[\iter]), \policy[\inneraction](\staterv[\iter])) \right]
\enspace ,
\end{align}
\noindent
where the expectation is with respect to $\staterv[0] \sim \initstates$ and $\staterv[\iter+1] \sim \trans (\cdot \mid \state[\iter], \policy[\outeraction](\staterv[\iter]), \policy[\inneraction](\staterv[\iter]))$.


In spite of multiple compelling applications, including autonomous driving \cite{fisac2015reach, leung2022learning}, reach-avoid problems in human-robot interaction \cite{bansal2017hamilton}, robust optimization in stochastic environments \cite{bertsimas2011theory}, and resource allocation over time \cite{goktas2022zero}, very little is known about computing recursive Stackelberg equilibria in zero-sum Markov Stackelberg games.
A
version of value iteration is known to converge in polynomial time under suitable assumptions \cite{goktas2022zero}, but this (planning) method becomes intractable in large or continuous state spaces.
To our knowledge, however, nothing is known about \emph{learning\/} Stackelberg equilibria from 
observed trajectories of play. 
In this paper, we introduce a novel class of zero-sum Markov Stackelberg games, namely convex-concave zero-sum Markov Stackelberg games, for which policy gradient-type methods can compute Stackelberg equilibria efficiently, and we show that this class of games can be used to model reach-avoid problems.

\amy{efficient number of iterations and number of trajectories, both}

\textbf{Contributions.}
\Cref{eq:min_max_stoch_stackelberg} reveals that the problem of computing Stackelberg equilibria in zero-sum Markov Stackelberg games is an instance of a
min-max optimization problems with \sdeni{coupling}{coupled} constraints. \amy{we need to say what coupling constraints are. something like:}
\sdeni{}{This optimization has ''coupled constraints'' as the set of actions available to the follower at each state is determined by the leader's choice.}
\citeauthor{goktas2021minmax} \cite{goktas2021minmax} studied coupled min-max optimization problems assuming an \emph{exact\/} first-order oracle, meaning one that returns a function's exact gradient and its exact value at any point in its domain.
As access to an exact oracle is an unrealistic assumption in Markov games, we develop a first-order method for solving these problems, assuming access to only a \emph{stochastic\/} first-order oracle, which returns noisy estimates of a function's gradient and its value at any point in its domain.
We show that our method converges in polynomial-time (\Cref{thm:min_max_convergence}) in a large class of coupled min-max optimization problems.


We then proceed to study zero-sum Markov
Stackelberg games, providing sufficient conditions on the action correspondence $\coupledactions: \states \times \outeractions \to \inneractions$, the rewards $\reward: \states \times \outeractions \times \inneractions \to \R$, and the transition probability function $\trans: \states \times \states \times \outeractions \times \inneractions \to \R_+$ to guarantee that the game is convex-concave (\Cref{lemma:concavity_params_follower} and \ref{lemma:convexity_params_leader}---\Cref{sec_app:proofs}). \amy{not sure what was intended here? \Cref{assum:convex_concave} or \Cref{assum:alt_convex_concave} or \Cref{assum:smooth})?}
Furthermore, we develop a policy gradient algorithm that provably converges to Stackelberg equilibrium in polynomial time in such convex-concave games (\Cref{thm:conv_pgda}) \amy{fix Cref}, the first reinforcement learning algorithm of this kind.
Finally, we prove that reach-avoid problems are naturally modeled in this way, and run experiments which show that the 
{the Stackelberg equilibrium policies learned by our method exhibit better safety and liveness properties than} their Nash counterparts.

\section{Preliminaries}

\textbf{Notation.}
All notation for variable types, e.g., vectors, should be clear from context; if any confusion arises, see \Cref{sec_app:prelims}.
We let $\simplex[n] = \{\x \in \R_+^n \mid \sum_{i = 1}^n x_i = 1 \}$ denote the unit simplex in $\R^n$, and $\simplex(A)$, the set of probability distributions on the set $A$.
We also define the support of any distribution $f \in \simplex(\calX)$ as $\supp(f) \doteq \left\{ \x \in \calX: f(\x) > 0 \right\}$.
We denote the orthogonal projection operator onto a set $C$ by $\project[C]$, i.e., $\project[C](\x) = \argmin_{\y \in C} \left\|\x - \y \right\|^2$.
Unless otherwise noted, we assume $\left\| \cdot \right\|$ is the Euclidean norm, $\left\| \cdot \right\|_2$.
\deni{I think removable now:}\amy{might still be used by Arjun in exp'ts.}
We denote by $\setindic[\calC](\x)$ the indicator function of a set $\calC$, with value 1 if $\x \in \calC$ and $0$ otherwise.
Given two vectors $\x, \y \in \R^n$, we write $\x \geq \y$ or $\x > \y$ to mean component-wise $\ge$ or $>$, respectively.
For any set $\calC$, we denote the diameter by $\diam (\calC) \doteq \max_{\c, \c^\prime \in \calC} \|\c - \c^\prime \|$. 

\textbf{Mathematical Concepts.}
\deni{SUBGRADIENT NOTATION!!}
%
Given $\calX \subset \R^\outerdim$, the function $\obj: \calX \to \R$ is said to be $\lipschitz[\obj]$-\mydef{Lipschitz-continuous} w.r.t.\ norm $\left\| \cdot \right\|$ iff $\forall \x_1, \x_2 \in \calX, \left\| \obj(\x_1) - \obj(\x_2) \right\| \leq \lipschitz[\obj] \left\| \x_1 - \x_2 \right\|$.
If the gradient of $\obj$ is $\lipschitz[\grad \obj]$-Lipschitz-continuous, we refer to $\obj$ as $\lipschitz[\grad \obj]$-\mydef{Lipschitz-smooth}. 
$\obj$ is \mydef{convex} (resp.\@ \mydef{concave}) iff for all $\lambda \in (0,1)$ and $\x, \x^\prime \in  \calX$, $\obj(\lambda \x + (1-\lambda)\x^\prime) \leq \text{(resp. $\geq$) } \lambda \obj(\x) + (1-\lambda)\obj(\x^\prime)$. 
$\obj$ is \mydef{log-convex} (resp.\@ \mydef{log-concave}) if $\log \circ \obj$ is convex (resp.\@ concave).
$\obj$ is $\mu$-\mydef{strongly convex (SC)} if $\obj(\x_1) \geq \obj(\x_2) + \left< \grad[\x] \obj(\x_2), \x_1 - \x_2 \right> + \nicefrac{\mu}{2} \left\| \x_1 - \x_1 \right\|^2$\sdeni{, and $\mu$-\mydef{strongly concave} if $-\obj$ is $\mu$-strongly convex}{}. 
For convenience, we say that an $l$-dimensional vector-valued function $\g: \calX \to \calY \subset \R^l$
is log-convex, convex, log-concave, or concave, respectively, if $g_k$ is log-convex, convex, log-concave, or concave, for all $k \in [l]$.
A correspondence $\coupledset: \calX \to \innerset$ is concave if for all $\lambda \in (0, 1)$ and $\outer, \outer[][][\prime] \in \calX$, $\coupledset(\lambda \outer + (1-\lambda) \outer[][][\prime]) \subseteq \lambda \coupledset(\outer) + (1-\lambda) \coupledset(\outer)$, assuming Minkowski summation 
\cite{czerwik2002functional, nikodem1989concave}.


We require notions of stochastic convexity related to stochastic dominance of probability distributions \cite{atakan2003valfunc}.
\sdeni{}{Given a choice set $\calW$,\amy{this is still bothering me. do we not need to assume a joint dist'n over $\calO$ and $\calW$ for this def'n to make sense?} \sdeni{a}{a joint} probability distribution $\trans \in \simplex (\calO \times \calW)$ over a continuous sample space $\calO$ and the choice set $\calW$ is said to be} \mydef{stochastically convex} (resp.\ \mydef{stochastically concave}) in $\w$ if for all continuous, bounded, and convex (resp.\ concave) functions $\statevalue: \calO \to \R$,  $\lambda \in (0,1)$, and $\w^\prime, \w^\dagger \in \calW$ s.t.\ $\mean[\w] = \lambda \w^\prime + (1-\lambda)  \w^\dagger$, it holds that $\Ex_{O \sim \trans(\cdot \mid \mean[\w])} \left[\statevalue(O)\right] \leq \lambda \Ex_{O \sim  \trans(\cdot \mid \w^\prime)} \left[\statevalue(O)\right] + (1-\lambda)\Ex_{O \sim \trans(\cdot \mid \w^\dagger)} \left[\statevalue(O)\right]$ (resp\@ $\geq$).
\section{Min-Max Optimization with Coupled Constraints}
\label{sec:basic_algos}

\deni{The constraint is a coupling constraint, and the optimization problem is a coupled optimization problem.}

\mydef{A min-max Stackelberg game}, denoted $(\outerset, \innerset, \obj, \constr)$, is a two-player, zero-sum game, where one player, called the leader, first commits to an action $\outer \in \outerset$ from her action space $\outerset \subset \R^\outerdim$, after which the second player, called the follower, takes an action $\inner \in \coupledset (\outer) \subset \innerset$ from a subset of of his action space $\innerset \subseteq \R^\innerdim$  determined by the action correspondence $\coupledset: \R^\outerdim \rightrightarrows \innerset$.
As is standard in the optimization literature, we assume throughout that the follower's action correspondence can be equivalently represented
via a \mydef{coupling constraint function} $\constr: \R^\outerdim \times \R^\innerdim \to \R^\numconstrs$ s.t.\ $\coupledset (\outer) \doteq \left\{ \inner \in \innerset \mid \constr(\outer, \inner) \geq \zeros \right\}$.
An action profile $(\outer, \inner) \in \outerset \times \innerset$ comprises actions for both players.
Once both players have taken their actions, the leader (resp.\ follower) receives a loss (resp.\ payoff) $\obj(\outer, \inner)$, defined by an \mydef{objective function} $\obj: \R^\outerdim \times \R^\innerdim \rightarrow \R$.  
We define the \mydef{marginal function} $\val (\outer) \doteq \max_{\inner \in \coupledset (\outer)} \obj(\outer, \inner)$, which, given an action for the leader, outputs her ensuing payoff, assuming the follower best responds.

The constraints in a min-max Stackelberg game are said to be \mydef{uncoupled} 
if $\coupledset (\outer) = \innerset$, for all $\outer \in \outerset$.
A min-max Stackelberg game is said to be \mydef{continuous} iff 1.~the objective function $\obj$ is continuous; 2.~the action spaces $\outerset$ and $\innerset$ are non-empty and compact; and 3.~the action correspondence $\coupledset$ is continuous, non-empty- and compact-valued.%
\footnote{See Theorem 5.9 and Example 5.10 of \citeauthor{rockafellar2009variational} \cite{rockafellar2009variational} for conditions on $\constr$ that guarantee the continuity of $\coupledset$ or Section 3 of \citeauthor{goktas2021minmax} \cite{goktas2021minmax}.}

\textbf{Stackelberg Equilibrium.}
The canonical solution concept for min-max Stackelberg games is the $(\varepsilon, \delta)$-\mydef{Stackelberg equilibrium (\sdeni{}{$\varepsilon-$SE or SE if $\varepsilon = 0$)}}, an action profile $\left(\outer[][][*], \inner[][][*] \right) \in \outerset \times \innerset$ 
s.t.\@ $\|\proj[\R_-][\constr(\outer[][][*], \inner[][][*])]\| \leq \delta$ and $\min_{\outer \in \outerset} \val (\outer[][][]) + \varepsilon  \geq \obj \left( \outer[][][*], \inner[][][*] \right) \geq  
\max_{\inner \in \coupledset (\outer[][][*])} \obj \left( \outer[][][*], \inner \right) - \delta$ \sdeni{}{for $\varepsilon, \delta \geq 0$}.%
\footnote{When the approximation Note that for $\delta > 0$, this definition of a $(\varepsilon, \delta)-$SE is more general than the one introduced by \citeauthor{goktas2021minmax} \cite{goktas2021minmax}, as it allows for the coupling constraint to be satisfied only approximately, which is necessary in this paper, as the coupling constraint can only be accessed via a stochastic oracle.} 
As a straightforward corollary of Theorem 3.2 of \citeauthor{goktas2021minmax} \cite{goktas2021minmax}, a SE is guaranteed to exist in continuous Stackelberg games. 
Moreover, the set of SE can be characterized as solutions to the following coupled min-max optimization problem:
$\min_{\outer \in \outerset} \max_{\inner \in \coupledset (\outer[][][])} \obj (\outer[][][], \inner[][][])$.

An alternative but weaker solution concept previously considered for min-max Stackelberg games \cite{tsaknakis2021minimax} is the $\varepsilon-$\mydef{generalized Nash equilibrium (\sdeni{}{$\varepsilon-$GNE or GNE if $\varepsilon = 0$)}}, i.e., $(\outer[][][*], \inner[][][*]) \in \outerset \times \coupledset (\outer[][][*])$ s.t. $\min_{\outer \in \outerset}  \obj \left(\outer, \inner[][][*] \right) + \varepsilon \geq \obj \left( \outer[][][*], \inner[][][*] \right) \geq 
\max_{\inner \in \coupledset (\outer[][][*])} \obj \left( \outer[][][*], \inner \right)-\varepsilon $   \sdeni{}{for some $\varepsilon \geq 0$}.
In general, the set of 0-GNE and SE need not intersect; as such GNE are not necessarily solutions of $\min_{\outer \in \outerset} \max_{\inner \in \coupledset (\outer[][][])} \obj (\outer[][][], \inner[][][])$ (see, Appendix A of \citet{goktas2021minmax}).
Furthermore, there is no 0-GNE whose value is less than the SE value of a game.
When a min-max Stackelberg game's constraints are uncoupled, a ($\varepsilon-$)GNE is called ($\varepsilon-)$\mydef{saddle point} or a ($\varepsilon-$)\mydef{Nash equilibrium}, and \emph{is\/} also an SE. 
Finally, a saddle point is guaranteed to exist \cite{sion1958general, neumann1928theorie} in continuous min-max Stackelberg games with uncoupled constraints, a convex-concave objective $\obj$, and convex action spaces $\outerset$ and $\innerset$, in which case such games have traditionally been referred to as convex-concave min-max (simultaneous-move) games \cite{abernethy2018faster}.

\textbf{Convex-Concave Games.}
A min-max Stackelberg game is said to be \mydef{convex-concave} if, in addition to being continuous, 1.~the marginal function $\val$ is convex in $\outer$; 2.~the objective function $\obj (\outer, \inner)$ is concave in $\inner$, for all $\outer \in \outerset$; 3.~the action spaces $\outerset$ and $\innerset$ are convex; and 4.~the action correspondence $\coupledset$ is convex-valued.
Note that this definition generalizes the definition of convex-concave min-max (simultaneous-move) game, because in such games, $\val$ is guaranteed to be convex in $\outer$ by Danskin's theorem \cite{danskin1966thm} since $\obj$ is convex in $\outer$.
Assuming access to an exact first-order oracle, an $(\varepsilon, \delta)$-SE of a convex-concave min-max Stackelberg game can be computed in polynomial time, assuming $\obj$ and $\constr$ are Lipschitz-smooth \cite{goktas2021minmax},
while the computation is NP-hard in continuous min-max Stackelberg games, even when $\outerset$ and $\innerset$ are convex, $\obj$ is convex-concave, and $\constr$ is affine \cite{tsaknakis2020concave}.

All the conditions that define a convex-concave Stackelberg game depend on the game primitives, namely $(\outerset, \innerset, \obj, \constr)$, and are well-understood (see, for instance, \citeauthor{rockafellar2009variational} \cite{rockafellar2009variational}), with the exception of the condition that the marginal function $\val$ be convex. 
While it is difficult to obtain necessary and sufficient conditions on the game primitives that ensure the convexity of $\val$, 
one 
possibility is to require $\obj$ to be convex in $(\outer, \inner)$ and $\coupledset$ to be concave.%
\footnote{See Section 2 of \citeauthor{nikodem1989concave} \cite{nikodem1989concave} and Chapter 36 of \citeauthor{czerwik2002functional} \cite{czerwik2002functional} for conditions on $\constr$ which guarantee that $\coupledset$ is concave and/or continuous and/or convex-valued.
Also, note that $\coupledset$ can be convex-\emph{valued}, i.e., for all $\outer \in \outerset$, $\coupledset (\outer)$ is convex, even if $\coupledset$ is concave \cite{nikodem1989concave}.}
\amy{if we have this possibility (the one just mentioned), why did G+G introduce new sufficient conditions? are these difficult to satisfy? difficult to verify? what? need a segue that explains the purpose of the G+G conditions:}\deni{Doesn't the first sentence in this paragraph not do that?}\amy{no. what is insufficient about the two more recently mentioned conditions, namely $\obj$ is convex in $(\outer, \inner)$ and $\coupledset$ is concave. why were G+G unhappy with these?}
The following sufficient conditions were introduced by \citeauthor{goktas2021minmax} \cite{goktas2021minmax}.

\begin{assumption}[Convex-Concave Assumptions]
\label{assum:convex_concave}
1.~The objective function $\obj (\outer, \inner)$ is convex in $(\outer, \inner)$ and concave in $\inner$, for all $\outer \in \outerset$%
;
2.~the action correspondence $\coupledset$ is concave; 
3.~the action spaces $\outerset$ and $\innerset$ are convex.
\end{assumption}

However, as these assumptions are only sufficient, they are not necessarily satisfied in all applications of convex-concave min-max Stackelberg game.
So the convexity of the marginal function must sometimes be established by other means.
To this end, we provide the following alternative set of sufficient assumptions, which we use to show that the reach-avoid Stackelberg game we study in \Cref{sec:reach-avoid} is convex-concave\footnote{\sdeni{}{We include detailed theorem statements, proofs, and additional results in the full version of the paper.}}:

\begin{assumption}[Alternative Convex-Concave Assumptions]
\label{assum:alt_convex_concave}
    1.~(Convex-concave objective) The objective $\obj (\outer, \inner)$ is convex in $\outer$, for all $\inner \in \innerset$, and concave in $\inner$, for all $\outer \in \outerset$;
    2.~(log-convex-concave coupling) the coupling constraint $\constr (\outer, \inner)$ is log-convex in $\outer$, for all $\inner \in \innerset$, and concave in $\inner$, for all $\outer \in \outerset$; and 
    3.~the action spaces $\outerset$ and $\innerset$ are convex.
\end{assumption}

\textbf{Computation.}
We now turn our attention to the computation of $(\varepsilon, \delta)$-SE in convex-concave min-max Stackelberg games, assuming access to an \mydef{unbiased first-order stochastic oracle} $(\objrv, \constrrv, \objdistrib, \constrdistrib)$ comprising random functions $\objrv: \R^\outerdim \times \R^\innerdim \times \objs \to \R$ and $\constrrv: \R^\outerdim \times \R^\innerdim \times \constrs \to \R^\numconstrs$ and sampling distributions $\objdistrib \in \simplex(\objs)$ and $\constrdistrib \in \simplex(\constrs)$ s.t.\@ $\Ex_{\obji \sim \objdistrib}[\objrv(\outer, \inner; \obji)] = \obj (\outer, \inner)$, $\Ex_{\constri \sim \constrdistrib} [\constrrv(\outer, \inner; \constri)] =  \constr (\outer, \inner)$, $\Ex_{\obji}[\gradobjrv[(\outer, \inner)] (\outer, \inner; \obji)] = \grad \obj (\outer, \inner)$, and $\Ex_{\constri} [\gradconstrrv[(\outer, \inner)] (\outer, \inner; \constri)] = \grad \constr (\outer, \inner)$.
%
The following assumptions are required for the convergence of our methods.
\begin{assumption}[Convergence Assumptions]
\label{assum:smooth}
1.~(Lipschitz game) $\obj$ and $\constr$ are Lipschitz-smooth;
2.~(concave representation) the coupling constraint $\constr (\outer, \inner)$ is concave in $\inner$ \deni{This edit is imo overfluous because the $\mapsto$ notation also describes what variable we are considering} \amy{true, but then let's be consistent. either use the mapsto notation or say concave in whatever. so far in this paper, we haven't usually used mapsto in favor of concave or convex in whatever. but we do in Section 4, so let's adopt that convention from the start.} \deni{Actually, here I reverted back to the other way of doing it because I am refering to the function with its name.} for all $\outer \in \outerset$;
3.~(Slater's condition) $\forall \outer \in \outerset, \exists \widehat{\inner} \in \innerset$ s.t.\ $\constr (\outer, \widehat{\inner}) > 0$; and 
4.~(stochastic oracle) there exists an unbiased first-order stochastic oracle $(\objrv, \constrrv, \objdistrib, \constrdistrib)$ with bounded variance s.t. for all $(\outer, \inner) \in \outerset \times \innerset$:
$\Ex [ \| \constrrv (\outer, \inner; \constri) \|^2 ] \leq \variance[\constr]$,
$\Ex [ \| \gradobjrv[(\outer, \inner)] (\outer, \inner; \obji) \|^2 ] \leq \variance[\grad \obj]$, and 
$\Ex [ \| \gradconstrrv[(\outer, \inner)] (\outer, \inner; \constri) \|^2 ] \leq \variance[\grad \constr]$, for some $\variance[\constr], \variance[\grad \obj], \variance[\grad \constr] \in \R_+$.
\end{assumption}
%
%
%
%
In the sequel, we rely on the following notation and definitions.
For any action $\outer \in \outerset$ of the leader, we can re-express the marginal function in terms of the \mydef{Lagrangian} $\lang (\inner, \langmult; \outer) \doteq \obj (\outer, \inner) + \left< \langmult, \constr (\outer, \inner) \right>$ (see, for instance, Section 5 of \citet{boyd2004convex}) as follows: $\val (\outer) = \max_{\inner \in \innerset} \min_{\langmult \in \R_+^\numconstrs}  \lang (\inner, \langmult; \outer)$.
Further, we define the follower's best-response correspondence $\innerset^*(\outer) \doteq \argmax_{\inner \in \innerset} \min_{\langmult \in \R_+^\numconstrs} \lang (\inner, \langmult; \outer)$, and for any action $\inner \in \innerset$ of the follower, the associated KKT multiplier correspondence $\langmults^*(\outer) \doteq \argmin_{\langmult \in \R_+^\numconstrs} \max_{\inner \in \innerset} \lang (\inner, \langmult; \outer)$. 
With these definitions in hand, under \Cref{assum:smooth}, we can build an unbiased first-order stochastic oracle $\langrv (\inner, \langmult; \outer, \obji, \constri) \doteq \objrv(\outer, \inner; \obji) + \left<\langmult, \constrrv(\outer, \inner; \constri) \right>$ for the Lagrangian $\lang$ s.t.\@ $\Ex_{(\obji, \constri)} [\langrv (\inner, \langmult; \outer, \obji, \constri)]$, where the expectation is taken over $(\obji, \constri) \sim \objdistrib \times \constrdistrib$.
of the stochastic first-order oracle $(\objrv, \constrrv, \objdistrib, \constrdistrib)$.





\textbf{Algorithms.} 
In convex-concave min-max Stackelberg games with uncoupled constraints, \samy{}{$(0, \delta)$-SE correspond exactly to the $\delta$-saddle points of the min-max optimization problem, i.e., \amy{INSERT DEFINITION HERE}.}\deni{I'm not sure I understand this sentence, it seems wrong also? Also not all SE are saddle points}
Assuming exact first-order oracle access, a natural approach to computing SE in these games is thus to simultaneously run projected gradient descent and projected gradient ascent on the objective function $\obj$ w.r.t.\@ $\outer \in \outerset$ and $\inner \in \innerset$,
i.e., for $t = 0, 1, 2, \hdots$, $(\outer[][][][\iter + 1], \inner[][][][\iter + 1]) \gets \proj[\outerset\times \innerset][(\outer[][][][\iter], \inner[][][][\iter]) + (-\grad[\outer] \obj, \grad[\inner] \obj)(\outer[][][][\iter], \inner[][][][\iter])]$, a method known under the names of \mydef{Arrow-Hurwicz-Uzawa}, \mydef{primal-dual}, and \mydef{gradient descent ascent (GDA)} \cite{arrow-hurwicz, arrow1958studies}.
Intuitively, 
any \sdeni{stationary}{fixed} point of \sdeni{these dynamics}{GDA}, i.e., $(\outer[][][*], \inner[][][*]) \in \outerset \times \innerset$ s.t. $\| \proj[\outerset\times \innerset][(\outer[][][*], \inner[][][*]) + (-\grad[\outer] \obj, \grad[\inner] \obj)(\outer[][][*], \inner[][][*])] \| \, = 0$, satisfies the necessary and sufficient optimality condition for an action profile to be a SE of a min-max Stackelberg game with uncoupled constraints.
More generally, in min-max Stackelberg games (with\emph{out\/} coupled constraints), the necessary and sufficient optimality condition for an action profile $(\outer[][][*], \inner[][][*]) \in \outerset \times \innerset$ to be a SE is $\| \proj[{\outerset \times \coupledset (\outer[][][*])}][(\outer[][][*], \inner[][][*]) + (-\grad[\outer] \val (\outer[][][*]), \grad[\inner] \obj (\outer[][][*], \inner[][][*]))] \| \, = 0$, where, for any leader action $\outerpoint \in \outerset$, $\grad[\outer] \val (\outerpoint) \doteq \lang (\inner[][][*] (\outerpoint), \langmult[][*] (\outerpoint); \outerpoint)$, for some $(\inner[][][*], \langmult[][*])(\outerpoint) \in \innerset^*(\outerpoint) \times \langmults^*(\outerpoint)$,\amy{i regret asking you to remove the hats above the x's. i think i may prefer them here, to define $\grad[\outer] \val (\outerpoint)$} by the subdifferential envelope theorem \cite{goktas2021minmax}.
The observation that any subgradient of $\grad[\outer] \val$ depends on the optimal Lagrange multipliers motivates a first-order method based on the gradient of the Lagrangian.

A min-max Stackelberg game can be seen as a three-player game $\min_{\outer \in \outerset} \max_{\inner \in \coupledset (\outer)} \obj (\outer, \inner) = \min_{\outer \in \outerset} \max_{\inner \in \innerset} \min_{\langmult \in \R_+^\numconstrs} \lang (\inner, \langmult; \outer)$, where the $\outer$-player moves first, and the $\inner$- and $\langmult$-players move second, simultaneously, because strong duality holds under \Cref{assum:smooth} (Slater's condition \cite{slater1959convex}) for the inner min-max optimization problem, i.e., $\max_{\inner \in \innerset} \min_{\langmult \in \R_+^\numconstrs} \lang (\inner, \langmult; \outer) = \min_{\langmult \in \R_+^\numconstrs} \max_{\inner \in \innerset}  \lang (\inner, \langmult; \outer)$. 
The problem of computing an SE can thus be reduced to the min-max optimization $\min_{(\outer, \langmult) \in \outerset \times \R_+^\numconstrs} \max_{\inner \in \innerset} \lang (\inner, \langmult; \outer)$, which we might hope to solve by running GDA on $\lang (\inner, \langmult; \outer)$ w.r.t.\@ $(\outer, \langmult)$ and $\inner$ over $\outerset \times \R^\numconstrs_+$ and $\innerset$, respectively.
However, although the Lagrangian $\lang (\inner, \langmult; \outer)$ is concave in $\inner$, for all $\outer \in \outerset$, it is not convex in $(\outer, \langmult)$, and its stationary points (i.e., points $(\inner[][][*], \langmult[][*]; \outer[][][*])$ s.t.\@ $\|\proj[{\innerset \times \R^\numconstrs_+ \times \outerset}][(\inner[][][*], \langmult[][*]; \outer[][][*]) + (\grad[\inner]\lang, -\grad[\langmult]\lang, -\grad[\outer]\lang) (\inner[][][*], \langmult[][*]; \outer[][][*])] \| \, = 0$)
do not coincide with SE even in simple convex-concave min-max Stackelberg games (see, Example 3.3 of \citet{goktas2022gda}).

As GDA fails in these games, \citeauthor{goktas2021minmax} \cite{goktas2021minmax} developed nested GDA, an \sdeni{alternative}{nested} first-order method for computing an $(\varepsilon, \delta)-$SE.
In the nested part of this algorithm, the \sdeni{}{the Lagrangian saddle point relaxation of the }inner maximization problem $\max_{\inner \in \innerset} \min_{\langmult \in \R_+^\numconstrs} \lang (\inner, \langmult; \outer)$ is solved by running GDA on $\lang$ w.r.t.\@ $\inner$ and $\langmult$ over constraint sets $\innerset$ and $\R^\numconstrs_+$ until convergence to a $\delta$-saddle point.
Then, exploiting the convexity of the marginal function $\val$, nested gradient descent runs a descent step on $\val$ w.r.t.\@ $\outer$, where, for any leader action $\outerpoint \in \outerset$, a subgradient $\grad[\outer] \val$ is approximated by $\widehat{\grad[\outer] \val} (\outerpoint) = \lang (\inner, \langmult; \outerpoint)$.
In this paper, we replace their exact first-order oracle with a stochastic one, the gradient descent step with a stochastic gradient descent (SGD) step, and GDA with stochastic GDA (SGDA), using in both cases the stochastic Lagrangian oracle $\langrv$.
We call our method nested stochastc gradient descent ascent (nested SGDA).

\begin{wrapfigure}{l}{0.64\textwidth}
    \vspace*{-0.8cm}
    \begin{minipage}{0.63\textwidth}
\begin{algorithm}[H]
\textbf{Inputs:} $\outerset, \innerset, \langmults , \objrv, \constrrv, \objdistrib, \constrdistrib, \outer[][][][0], \numiters[\outer], \numiters[\inner], (\learnrate[\outer][\iter])_\iter, (\learnrate[\inner][\iter])_{\iter}, \delta$ \\
\textbf{Outputs:} $(\outer[][][][\iter], \inner[][][][\iter], \langmult[][][\iter])_{\iter = 0}^{\numiters[\outer]-1}$ 

\begin{algorithmic}[1]

\caption{Saddle-Point-Oracle \sdeni{Stochastic Gradient Descent}{SGD}/Nested \sdeni{Stochastic GDA}{SGDA}\label{alg:combined}}
\For{$\iter = 0, \hdots, \numiters[\outer] - 1$}
    \If{\mydef{Saddle-Point-Oracle SGD}}
    \State Find $(\inner[][][][\iter], \langmult[][][\iter]) \in \innerset \times \R^\numconstrs_+$ s.t. 
    \State $\max\limits_{\inner \in \innerset} \lang (\inner[][][][\iter], \langmult; \outer[][][][\iter]) - \min\limits_{\langmult \in \R_+^\numconstrs} \lang (\inner, \langmult[][][\iter]; \outer[][][][\iter])\leq \delta$,
    \EndIf
    \If{\mydef{Nested SGDA}}
        \State $(\inner[][][][\iter], \langmult[][][\iter]) \gets$ Initialize arbitrarily
        \For{$s = 0, \hdots, \numiters[\inner] - 1$}
            
            \State Sample $\obji \sim \objdistrib, \constri \sim \constrdistrib$
            
            \State $\inner[][][][\iter] \! \! \gets  \! \! \project[{\innerset}]\! \!\left[\inner[][][][\iter] \! \! + \learnrate[\inner][\iter] \gradlangrv[\inner] (\inner[][][][\iter]\!\!, \langmult[][][\iter]; \outer[][][][\iter]\!\!, \obji, \constri) \right]$
            
            \State  $\langmult[][][\iter] \! \!\gets  \! \!\project[\langmults]\! \!\left[\langmult[][][\iter] \! \! -  
            \learnrate[\inner][\iter] \gradlangrv[\langmult] (\inner[][][][\iter]\!\!, \langmult[][][\iter]; \outer[][][][\iter]\!\!, \obji, \constri) \right]$

                
        
    \EndFor
    \EndIf
    \State Sample $\obji \sim \objdistrib, \constri \sim \constrdistrib$
    
    
    \State $\outer[][][][\iter + 1] \! \! \gets \! \! \project[\outerset]\! \!\left[\outer[][][][\iter]\! - \!  
    \learnrate[\outer][\iter]\gradlangrv[\outer] (\inner[][][][\iter], \langmult[][][\iter]; \outer[][][][\iter], \obji, \constri) \right]$
\EndFor

\State \Return $(\outer[][][][\iter], \inner[][][][\iter], \langmult[][][\iter])_{\iter = 0}^{\numiters[\outer]-1}$ 
\end{algorithmic}
\end{algorithm}
    \end{minipage}
    \vspace*{-0.3cm}
\end{wrapfigure}

We begin by presenting \mydef{saddle-point-oracle stochastic gradient descent algorithm} (saddle-point-oracle SGD---\Cref{alg:combined}), whose analysis we build on to develop our primary contribution, nested SGDA.
Following \citeauthor{goktas2021minmax} \cite{goktas2021minmax}, saddle-point-oracle SGD runs SGD on $\val$, assuming access to an oracle, which, for any leader $\outer \in \outerset$, returns a $\delta$-saddle point of $(\inner, \langmult) \mapsto \lang (\inner, \langmult; \outer)$.
But, whereas they assume access to a $\delta$-oracle that takes as input any action of the leader $\outer \in \outerset$ and returns a tuple $(\tilde{\inner} (\outer),\tilde{\langmult} (\outer))$ consisting of
1.~a $\delta$-best-response for the follower, i.e., $\tilde{\inner} (\outer) \in 
\coupledset (\outer)$ s.t.\@ $\obj (\outer, \tilde{\inner} (\outer)) + \delta \geq \max_{\inner[][][\prime] \in \coupledset (\outer)} \obj (\outer, \inner[][][\prime])$, and 2.~an associated Lagrange multiplier $\tilde{\langmult} (\outer) \in \langmults^*(\outer)$, we assume access to a $\delta$-saddle-point oracle.
%
Our second algorithm, \mydef{nested stochastic gradient descent ascent} (nested SGDA---\Cref{alg:combined}), follows the same logic as saddle-point-oracle SGD, but  implements the saddle-point oracle by running \sdeni{stochastic GDA}{SGDA.}
Nested SGDA generalizes \citet{goktas2021minmax}'s nested GDA algorithm from a setting with an exact first-order oracle to a stochastic one. 
The following theorem establishes conditions under which both of our algorithms converge to an $(\varepsilon + \delta, \delta)$-Stackelberg equilibrium in a polynomial number of oracle evaluations in the parameters of the problem, $\nicefrac{1}{\varepsilon}$, $\nicefrac{1}{\delta}$.%

\begin{restatable}[]{theorem}{thmminmaxconvergence}
\label{thm:min_max_convergence}
Let $(\outerset, \innerset, \obj, \constr)$ be a convex-concave min-max Stackelberg game for which \Cref{assum:smooth} holds.
For any $\varepsilon, \delta \geq 0$, if nested \sdeni{stochastic GDA}{SGDA} (resp.\ saddle-point-oracle \sdeni{stochastic gradient descent}{SGD}) is run with inputs that satisfy for all $\iter \in \N_+$, $\learnrate[\outer][\iter], \learnrate[\outer][\iter] \in \Theta\left (\nicefrac{1}{\sqrt{\iter + 1}} \right)$, and outputs $(\outer[][][][\iter], \inner[][][][\iter], \langmult[][][\iter])_{\iter = 0}^{\numiters[\outer]-1}$, then, in expectation over all runs of the algorithm, the average action profile $(\mean[{\outer[][][][]}], \mean[{\inner[][][][]}])$
is an $(\varepsilon+\delta, \delta)$-Stackelberg equilibrium after $\tilde{O} (\nicefrac{1}{\varepsilon^2 \delta^2})$ (resp.\@ $\tilde{O} (\nicefrac{1}{\varepsilon^2})$)  gradient evaluations.
%
%
If, in addition, $\val$ is strongly-convex (e.g., \Cref{assum:convex_concave} holds and $\obj (\outer, \inner)$ is $\scparam[\outer]$-strongly-convex in $\outer$), then
$(\mean[{\outer[][][][]}], \mean[{\inner[][][][]}])$ 
is an $(\varepsilon + \delta, \delta)$-Stackelberg equilibrium after $\tilde{O} (\nicefrac{1}{\varepsilon \delta^2})$ (resp.\@ $\tilde{O} (\nicefrac{1}{\varepsilon})$) gradient evaluations.
In both cases, $\mean[{\outer[][][]}] \doteq \frac{\sum_{\iter = 0}^{\numiters[\outer] -1}  \learnrate[\outer][\iter]\outer[][][][\iter]}{\sum_{\iter = 0}^{\numiters[\outer] - 1}  \learnrate[\outer][\iter]}$ and $\mean[{\inner[][][]}] \doteq \frac{\sum_{\iter = 0}^{\numiters[\outer] -1}  \learnrate[\outer][\iter]\inner[][][][\iter]}{\sum_{\iter = 0}^{\numiters[\outer] -1}  \learnrate[\inner][\iter]}$.

\end{restatable}

\section{Policy Gradient in Convex-Concave Zero-Sum Markov Stackelberg Games}\label{sec:policy_grad}

In this section, we formulate the computation of Stackelberg equilibrium in zero-sum Markov Stackelberg games as a min-max optimization problem, which enables us to derive a policy gradient method based on our nested SGDA algorithm. 
\sdeni{Additionally, we derive sufficient conditions on any continuous state and action zero-sum Markov Stackelberg game such that the ensuing min-max optimization problem satisfies the assumptions of \Cref{thm:min_max_convergence}, thereby obtaining polynomial-time convergence results for nested SGDA in a new class of games, 
assuming \mydef{first-order stochastic oracle} access to the gradient of the reward and transition probability functions of the game.}{}

We consider zero-sum Markov Stackelberg games $\game \doteq (\states, \outeractions, \inneractions, \initstates, \reward, \constr, \trans, \discount)$ with action spaces $\outeractions \subset \R^\outerdim$ and $\inneractions \subset \R^\innerdim$ for the leader and follower, respectively, 
where the follower's actions are constrained by the leader's via vector-valued state-dependent coupling constraints $\constr$
that define a correspondence $\coupledactions(\state, \outeraction) \doteq \{\inneraction \in \inneractions \mid \constr(\state, \outeraction, \inneraction) \geq \zeros\}$.
We assume without loss of generality that the initial state distribution $\initstates \in \simplex(\states)$ has full support over the state space,%
\footnote{This assumption is without loss of generality, as if it does not hold, one can replace the initial state distribution with an arbitrary initial state distribution with full support over the state space without affecting any of our results, because recursive Stackelberg equilibria are independent of the initial state distribution.}
i.e., $\supp(\initstates) = \states$.
\sdeni{}{A \mydef{Markov} policy for the leader (resp. follower)---hereafter policy for short---is one that is  history independent, and thus a mapping from states to actions $\policy[\outeraction]: \states \to \outeractions$ (resp. $\policy[\inneraction]: \states \to \outeractions$).}\amy{is a stationary policy really a thing? a policy is a mapping from histories to actions. a Markov policy (which is a thing) happens to depend only on the current state}
\sdeni{While in principle a player's policy can be history-dependent, our focus is on Markov policies, which are mapping from states to actions.}{}
Overloading notation, we define the correspondence of feasible policies $\coupledactions (\policy[\outeraction]) \doteq \{ \policy[\inneraction] : \states \to \inneractions \mid 
\constr (\state, \policy[\outeraction](\state), \policy[\inneraction](\state) \geq \zeros, \mbox{ for all } \state \in \states 
\}$. 
As the initial state distribution has full support over the state space, the follower's feasible policy correspondence can be represented as $\coupledactions (\policy[\outeraction]) = \{ \policy[\inneraction] : \states \to \inneractions \mid \Ex_{\staterv \sim \rho} [\proj[\R_-][\constr (\staterv, \policy(\staterv))]] \geq \zeros\}$, which, overloading notation, we denote by $\constr(\policy[\outeraction], \policy[\inneraction]) \doteq \Ex_{\staterv \sim \rho} [\proj[\R_-][\constr (\staterv, \policy(\staterv))]]$. \sdeni{}{A zero-sum Markov Stackelberg game is said to be \mydef{continuous} if 1.~for all states $\state \in \states$, the reward function $(\outeraction, \inneraction) \mapsto \reward (\state, \outeraction, \inneraction)$
is continuous and bounded, i.e., $\left\| \reward \right\|_{\infty} \leq \rewardbound < \infty$, for some $\rewardbound \in \R_{+}$;
2.~the action spaces $\outeractions$ and $\inneractions$ are non-empty and compact; 
and
3.~for all states $\state \in \states$, $\coupledactions (\state, \cdot)$ is continuous, non-empty, and compact-valued.}

%

A \mydef{history} $\hist[][][] \in (\states \times \outeractions \times \inneractions)^\numiters$ of length $\numiters$ is a sequence of state-action tuples $\hist[][][] = (\state[\iter], \outeraction[][][][\iter], \inneraction[][][][\iter])_{\iter = 0}^{\numiters-1}$.
Given a policy profile $\policy$ and a history of play $\hist[][][]$ of length $\numiters$, we define the \mydef{discounted history distribution} as
%
    $
    \histdistrib[][\policy][\numiters] (\hist[][][]) = \initstates (\state[0]) \prod_{\iter = 0}^{\numiters-1} \discount^\iter \trans (\state[\iter +1] \mid \state[\iter], \outeraction[][][][\iter], \inneraction[][][][\iter]) \mathbbm{1}_{\policy (\state[\iter])} (\outeraction[][][][\iter], \inneraction[][][][\iter]) 
    .
    $
%
Define the set of all realizable trajectories $\hists[\policy][\numiters]$ of length $\numiters$ under policy $\policy$ as $ \hists[\policy][\numiters] \doteq \supp(\histdistrib[][\policy][\numiters])$, i.e., the set of all histories that occur with non-zero probability.
For convenience, we denote by
$\histdistrib[][\policy] \doteq \histdistrib[][\policy][\infty]$, and by $\histrv[][] = \left(\staterv[\iter], \outeractionrv[\iter], \inneractionrv[\iter] \right)_\iter$ any randomly sampled history from $\histdistrib[][\policy]$.
Finally, we define the \mydef{discounted state-visitation distribution} under any initial state distribution $\initstates$ as {$\statedist[{\initstates}][\policy] (\state) = \sum_{\iter = 0}^\infty\sum_{\hist \in \hists[\policy][\iter]: \staterv[\iter] = \state}  \initstates(\state[0])\prod_{k = 1}^\iter \discount^k \trans (\state[k] \mid \state[k-1], \outeraction[][][][k-1], \inneraction[][][][k-1])$.



\amy{language: i think for one-shot we use reward, and for long--term (discounted cumulative), payoffs}

Given a zero-sum Markov Stackelberg game $\game$ and a policy profile $\policy \doteq (\policy[\outeraction], \policy[\inneraction]) \in \outeractions^\states \times \inneractions^\states$, the \mydef{state-value function} $\statevalue[][\policy]: \states \to \R$ and the \mydef{action-value function} $\actionvalue[][\policy]: \states \times \outeractions \times \inneractions \to \R$ are defined in terms of $\histdistrib[][\policy]$ as
    $\statevalue[][{\policy}] (\state) \doteq \mathop{\Ex}_{\histrv \sim \histdistrib[][\policy]} \left[ \sum_{\iter = 0}^\infty
    \discount^\iter \reward \left( \staterv[\iter], \outeractionrv[\iter], \inneractionrv[\iter] \right) 
    \mid  
    \staterv[0] = \state
    \right]$
and
    $\actionvalue[][{\policy}](\state, \outeraction, \inneraction) \doteq 
    \mathop{\Ex}_{\histrv \sim \histdistrib[][\policy]} \left[ \sum_{\iter = 0}^\infty
    \discount^\iter \reward \left( \staterv[\iter], \outeractionrv[\iter], \inneractionrv[\iter] \right) 
    \mid  
    \staterv[0] = \state,
    \outeractionrv[0] = \outeraction,
    \inneractionrv[0] = \inneraction
    \right]$, respectively.
%
%
%
The \mydef{cumulative payoff function} $\cumulutil: \outeractions^\states \times \inneractions^\states \to \R$ is then defined in terms of the state-value function as $\cumulutil (\policy[\outeraction], \policy[\inneraction]) \doteq \Ex_{\staterv \sim \initstates} \left[ \statevalue[][{\policy}] (\staterv)\right]$, i.e., the total expected loss (resp.\ gain) of the leader (resp.\ follower). Finally, for any leader policy $\policy[\outeraction]: \states \to \outeractions$, we also define the \mydef{marginal state-value function} $\marginalstatefunc[][{\policy[\outeraction]}] (\state) \doteq \max_{\policy[\inneraction] \in \coupledactions(\policy[\outeraction])
} \statevalue[{\policy[\outeraction], \policy[\inneraction]}](\state)$,
which is the leader's payoff from state $\state$ onward, assuming the follower best responds with a feasible policy, as well as the (expected) \mydef{marginal function} $\marginalfunc(\policy[\outeraction]) \doteq \Ex_{\staterv \sim \initstates} \left[ \marginalstatefunc[][{\policy[\outeraction]}](\staterv) \right]$.

%

\textbf{Recursive Stackelberg Equilibrium.}
A policy profile $\policy[][][*] \doteq (\policy[\outeraction][][*], \policy[\inneraction][][*]) \in \outeractions^\states \times \inneractions^\states$ is called an $(\varepsilon, \delta)$-\mydef{recursive (or Markov perfect) Stackelberg equilibrium} iff $\ \| \constr(\policy[][][*])] \| \ \leq \delta$ and
$\max_{\policy[\inneraction] \in \coupledactions (\policy[\outeraction])} \statevalue[][{\policy[\outeraction][][*], \policy[\inneraction]}] (\state) - \delta \leq \statevalue[][{\policy[\outeraction][][*], \policy[\inneraction][][*]}](\state) \leq \min_{\policy[\outeraction] \in \outeractions^\states} \marginalstatefunc[][{\policy[\outeraction]}] (\state) + \varepsilon = \min_{\policy[\outeraction] \in \outeractions^\states} \max_{\policy[\inneraction] \in \coupledactions (\policy[\outeraction])} \statevalue[][{\policy[\outeraction], \policy[\inneraction]}] (\state) + \varepsilon$, for all states $\state \in \states$.
\sdeni{}{A recursive Stackelberg equilibrium is guaranteed to exist in continuous zero-sum Markov Stackelberg games \cite{goktas2022zero}.}

As our plan is to apply our nested SGDA algorithm to compute recursive Stackelberg equilibria, we begin by showing that zero-sum Markov Stackelberg games are an instance of min-max Stackelberg games.
Assume parametric policy classes for the leader and follower, respectively, namely $\policies[\outerset] \doteq \{ \policy[\outer]: \states \to \outeractions \mid \outer \in \outerset \} \subseteq \outeractions^\states
$ and $\policies[\innerset] \doteq \{ \policy[\inner]: \states \to \inneractions \mid \inner \in \innerset \} \subseteq \inneractions^\states 
$, for \sdeni{}{non-empty and compact} parameter spaces $\outerset \subset \R^\numparams$ and $\innerset \subset \R^\numparams$.
Using these parameterizations, we redefine
    $\statevalue[][{(\outer, \inner)}] \doteq \statevalue[][{(\policy[\outer], \policy[\inner])}]$,
    $\actionvalue[][{(\outer, \inner)}] \doteq \actionvalue[][{(\policy[\outer], \policy[\inner])}]$,
    $\cumulutil(\outer, \inner) \doteq \cumulutil(\policy[\outer], \policy[\inner])$,
    etc.,
and thus restate the problem of computing a recursive Stackelberg equilibrium as finding $(\outer, \inner) \in \outerset \times \innerset$ that solves 
    $\min_{\outer \in \outerset} \max_{\inner \in \coupledactions (\outer)
    } \statevalue[][{(\outer,\inner)}] (\state)$, for all states $\state \in \states$. 
In fact, we solve this problem in expectation over the initial state distribution $\initstates$, i.e., $\min_{\outer \in \outerset} \max_{\inner \in \coupledactions (\outer)} \cumulutil(\outer, \inner) = \min_{\outer \in \outerset} \max_{\inner \in \innerset : \constr(\outer, \inner) \geq 0} \cumulutil(\outer, \inner)$ 

\textbf{Convex-Concave Markov Games.}
As we have shown (\Cref{thm:min_max_convergence}), Stackelberg equilibria can be computed in polynomial time in convex-concave min-max Stackelberg games, assuming access to an unbiased first order-stochastic oracle.
We now define an analogous class of Markov Stackelberg games, namely \mydef{convex-concave zero-sum Markov Stackelberg games}---continuous zero-sum Markov Stackelberg games for which 1.~the marginal function $\outer \mapsto \marginalfunc(\outer)$ is convex, 2.~the cumulative payoff function $\inner \mapsto \cumulutil (\outer, \inner)$ is concave\sdeni{}{, 3.~the policy spaces $\outerset$ and $\innerset$ are convex, 3.~ $\outer \mapsto \coupledset(\outer)$ is convex-valued} \amy{let's try to be consistent! i like the mapsto notation for this section for sure.} for all $\outer \in \outerset$---for which a similar result holds. 
%

In \Cref{sec_app:proofs}, we provide a sufficient characterization of convex-concave zero-sum Markov Stackelberg games (\Cref{lemma:concavity_params_follower} and \ref{lemma:convexity_params_leader}). From these characterizations we can conclude that under a bilinear parameterization, sufficient conditions for a zero-sum Markov Stackelberg game to be convex-concave are for the 1.~reward (resp. transition probability) function to be affine (resp. stochastically affine) in the follower's action, 2.~the reward (resp. transition probability) function to be convex (resp. stochastically convex) in the follower's action, and 3.~the follower's action correspondence to be concave.


\textbf{Computation.}
We now turn our attention to the computation of recursive Stackelberg equilibrium in convex-concave zero-sum Markov Stackelberg games.
Mirroring the steps by which policy gradient has been show to converge in other settings \cite{daskalakis2020independent}, we
first define an unbiased first-order stochastic oracle for zero-sum Markov-Stackelberg games, given access to an unbiased first-order stochastic oracle for the reward and probability transition function; and we then establish convergence of nested SGDA in this setting by invoking \Cref{thm:min_max_convergence}.
\sdeni{Note that our estimator generalizes the classic REINFORCE estimator \cite{williams1992simple} from discrete state and action spaces to continuous.}{}

Stochastic nested GDA relies on an unbiased first-order stochastic oracle $(\objrv, \constrrv, \objdistrib, \constrdistrib)$, which we can use to obtain unbiased first-order stochastic estimates of $\cumulutil$ and $\constr$. 
Setting, $\constrrv (\outer, \inner; \state) \doteq \proj[\R_-][\constr(\state, \policy[\outer](\state), \policy[\inner](\state))]$ and $\constrdistrib(\state) \doteq \initstates(\state)$, we obtain an unbiased first-order stochastic oracle for the constraints $\constr$.
%
%
Regarding, the estimation of $\grad \cumulutil$, the deterministic policy gradient theorem \cite{silver2014deterministic} provides sufficient conditions (\Cref{assum:smooth_stoch}) for the state-value function of MDPs to be differentiable. 
A straightforward generalization of this theorem holds for differentiating $\cumulutil$ w.r.t.\@ $\outer$ in zero-sum Markov Stackelberg games, so that $\grad[\outer] \cumulutil (\outerpoint, \innerpoint) = \mathop{\Ex}_{\substack{\staterv \sim \statedist[\initstates][{(\outerpoint, \innerpoint)}]}} \left[ \grad[\outeraction] \actionvalue[][{(\outerpoint, \innerpoint)}](\staterv, \policy[\outerpoint](\staterv), \policy[\innerpoint](\staterv))  \grad[\outer] \policy[\outerpoint](\staterv) \right]$. 
An analogous result holds for the gradient w.r.t.\@ $\inner$.
\begin{assumption}[Convergence Assumptions]\label{assum:smooth_stoch}
1.~The parameter space $\outerset$ and $\innerset$ are non-empty, compact, and convex;
2.~(Slater's condition)
for all $\state \in \states, \outeraction \in \outeractions$, there exists $\widehat{\inneraction} \in \inneractions$ s.t.\ $\constr(\state, \outeraction, \widehat{\inneraction}) > 0$; \amy{$\numconstr$ is not quantified} and
3.~$\trans$, $\reward$, and $\constr$
are twice continuously differentiable.
%
\end{assumption}
%
To invoke the deterministic policy gradient theorem requires access to two gradients.
The gradient $\grad[\outer] \policy[\outer] (\staterv)$ of the policy w.r.t.\@ its parameters is available by design in all differentiable parameterizations of interest. 
The gradient $\grad[(\outeraction, \inneraction)] \actionvalue[\outer, \inner]$ of the state-action value function w.r.t\@ both players' actions can be computed via the chain rule, since our oracle affords us access to the gradients of both the reward and probability transition functions: 
$\grad[(\outeraction, \inneraction)] \actionvalue[\outer, \inner](\state, \outeraction, \inneraction) = \Ex_{\hist} [\gradactionvalue[](\state, \outeraction, \inneraction, \outer, \inner; \hist)]$.
This gradient involves an expectation over all possible histories; nonetheless, we can estimate it from history samples $\histrv \sim \histdistrib[][(\outer, \inner)] \mid \staterv[0] = \state, \outeractionrv[0] = \outeraction, \inneractionrv[0] = \inneraction$: i.e., we choose as our estimator $\gradobjrv (\outer, \inner; \hist) \doteq \gradactionvalue[](\state[0], \outeraction[][][][0], \inneraction[][][][0], \outer, \inner ;\hist) \grad (\policy[\outer], \policy[\inner])(\state[0])$. 
Since the state and action spaces are compact, and the reward and transition probability functions are continuously differentiable, the gradient of the state-value \amy{double checking that you mean state and not action-value? since the estimator is estimating Q not V?} function is bounded over its domain; hence the variance of our estimator must likewise be bounded, i.e., $\variance[\grad \obj] \doteq  \|\gradobjrv (\outer, \inner; \hist) \|_\infty^2 \geq \Ex_{\hist}[\|\gradobjrv (\outer, \inner; \hist) \|^2]$.  
With all of this machinery in place, we can now extend nested SGDA to 
compute recursive Stackelberg equilibria in zero-sum Markov Stackelberg games  (\Cref{alg:nested_pgda}; \Cref{sec:algos}).

\begin{restatable}{theorem}{thmconvpgda}
\label{thm:conv_pgda}
Let $\game$ be a \sdeni{}{convex-concave} zero-sum Markov Stackelberg game \sdeni{satisfying the assumptions of}{(e.g., } \Cref{lemma:concavity_params_follower} and \Cref{lemma:convexity_params_leader} \sdeni{}{holds)}. 
Under \Cref{assum:smooth_stoch}, for any $\varepsilon, \delta \geq 0$, if nested policy gradient descent ascent (\Cref{alg:nested_pgda})
is run with inputs that satisfy for all $\iter \in \N_+$, $\learnrate[\outer][\iter], \learnrate[\outer][\iter] \in \Theta\left(\nicefrac{1}{\sqrt{\iter + 1}} \right)$, and outputs $(\outer[][][][\iter], \inner[][][][\iter], \langmult[][][\iter])_{\iter = 0}^{\numiters[\outer]-1}$, then the policy profile $(\policy[{\mean[{\outer[][][]}]}], \policy[{\mean[{\inner[][][]}]}])$ is an $(\varepsilon+\delta, \delta)$ recursive Stackelberg equilibrium after $\tilde{O}(\nicefrac{1}{\varepsilon^2 \delta^2})$ gradient evaluations, where
$\mean[{\outer[][][]}] \doteq \frac{\sum_{\iter = 0}^{\numiters[\outer] -1}  \learnrate[\outer][\iter]\outer[][][][\iter]}{\sum_{\iter = 1}^{\numiters[\outer]}  \learnrate[\outer][\iter]}$ and 
$\mean[{\inner[][][]}] \doteq \frac{\sum_{\iter = 0}^{\numiters[\inner] -1}  \learnrate[\outer][\iter]\inner[][][][\iter]}{\sum_{\iter = 1}^{\numiters[\outer]}  \learnrate[\inner][\iter]}$.
\end{restatable}

\section{Application: Reach-Avoid Problems}
\label{sec:reach-avoid}

\amy{
\begin{itemize}
\item we need convexity of the marginal function in $\outer$. sufficient condition: concave correspondence. in the discrete-action RA games, we are guaranteed a concave correspondence, b/c the target set is concave -- but why is it concave?
\item we also need concavity of the objective function in $\inner$. we cannot achieve concavity. but we can achieve incavity.
\item our convergence theorems should generalize to convex-incave games (though we do not plan to work through this generalization). instead, we run experiments.
\end{itemize}
}


\sarjun{Reach-avoid problems are closely related to persuit-evasion differential games. In persuit-evasion, one player seeks to collide with another fleeing player \cite{isaacs1954differential}.}{}

\samy{In the reach-avoid problem one agent, the protagonist, aims to reach a target while the other agent, called the antagonist tries to capture the protagonist.}{} \sarjun{in which the agent, now called the protagonist, aims to reach a target, while another agent, called the antagonist, tries to 
capture the protagonist.}{} 

In this section, we endeavor to apply our algorithms to a real-world application, namely reach-avoid problems.
In a reach-avoid problem (e.g., \cite{fisac2015reach, gao2007reachability}), an agent seeks to reach one of a set of targets---achieve \emph{liveness}---while avoiding 
obstacles along the way---ensuring \emph{safety}. 
Reach-avoid problems have myriad applications, including network consensus problems \cite{khanafer2013robust}, motion planning \cite{chen2014multiplayer, karaman2011sampling}, pursuit-evasion games \cite{flynn1974lion, lewin1986lion}, autonomous driving \cite{leung2022learning}, and path planning \cite{zhou2018efficient}, to name a few.
\amy{what about trying to rescue hostages from terrorists?}

The obstacles in a reach-avoid problem are not necessary stationary; they may move, either randomly or deliberately, around the environment.
When the obstacles' movement is random, the problem can be modeled and solved as an MDP.
On the other hand, when their movement is deliberate, so that they are more like a rational opponent than a stochastic process, the problem is naturally modeled as a zero-sum game, where the agent---the protagonist---aims to reach its target, while an antagonist (representing the obstacles) seeks to prevent the protagonist from doing so.
Past work has modeled these games as simultaneous-move (e.g., \cite{fisac2015reach}) \amy{what about \cite{gao2007reachability}? do they also model the problem as a game?}, imposing the constraint that the agent cannot collide with an obstacle via huge negative rewards.

We instead model this hard constraint properly, using the framework of zero-sum Markov Stackelberg games, with the leader as the antagonist, whose movement's impose constraints on the move of the follower, the protagonist. We then use nested policy GDA to compute Stackelberg equilibria and simultaneous SGDA to compute \sdeni{Nash equilibria}{GNE}, and show that the protagonist learns stronger policies in the sequential (i.e., Stackelberg) game than in the simultaneous.
Traditionally, reach-avoid games have been formulated as continuous-time, i.e., differential, games (e.g., \cite{friedman2013differential}).
As practical solutions to these games are ultimately implemented in discrete time, 
we directly model the problem as a \emph{discrete-time\/} discounted infinite-horizon zero-sum Markov game.

\amy{so should we just be outputting the values of the two inner maximization problems, instead of simulating the policies and tallying wins?: ``While it is a theorem that the antagonist's value at a Stackelberg equilibrium is better than its value at a (generalized) Nash equilibrium, we further show experimentally that the protagonist's value is also better.''} 

A (discrete-time discounted infinite-horizon continuous state and action) \mydef{reach-avoid game}
$(\numstates, \states, \targetstates, \safestates, \outeractions, \inneractions,  \initstates, \reward, \detertrans)$ comprises two players, the \mydef{antagonist} (or $\outeraction$-player) and the \mydef{protagonist} (or $\inneraction$-player), each of whom occupies a state $\state[][\outeraction], \state[][\inneraction] \in \states$ in a state space $\states \subset \R^\numstates$, for some $\numstates \in \N$.
The protagonist's goal is to find a path through the safe set $\safestates \subset \states \times \states$ that reaches a state in the target set $\targetstates \subset \safestates$, while steering clear of the avoid set
$\avoidstates = \states \times \states \setminus \safestates$. 
This safe and avoid set formulation is intended to represent capture constraints, which have been the focus of the reach-avoid literature \cite{zhou2018efficient}.

Initially, the players occupy some state $\state[0] \sim \initstates \in \simplex(\safestates) $ drawn from an initial joint distribution $\initstates$ over all states, excluding the target and avoid sets.
At each subsequent time-step $\iter \in \N_+$, the antagonist (resp.\@ protagonist) chooses $\outeraction[][][][\iter] \in \outeractions$ (resp.\@ $\inneraction[][][][\iter] \in \inneractions$) from a set of possible directions $\outeractions \subseteq \R^\numstates$ (resp.\@ $\inneractions \subseteq \R^\numstates$) in which to move.
After both the antagonist and the protagonist move,
they receive respective rewards $-\reward (\state[\iter], \outeraction[][][][\iter], \inneraction[][][][\iter])$ and $\reward (\state[\iter], \outeraction[][][][\iter], \inneraction[][][][\iter])$.
Then, either the game ends, with probability $1-\discount$, for some discount rate $\discount \in (0, 1)$, or
the players move to a new state $\state[\iter + 1] \doteq \detertrans (\state[\iter], \outeraction, \inneraction) = \left(\detertrans[\outeraction] ( \state[\iter][\outeraction], \outeraction), \detertrans[\inneraction] (\state[\iter][\inneraction], \inneraction) \right)$, as determined by their respective displacement functions $\detertrans[\outeraction]: \states \times \outeractions \to \states$ and $\detertrans[\inneraction]: \states \times \inneractions \to \states$. 
%


We define the feasible action correspondence $\coupledactions(\state, \outeraction) \doteq \{ \inneraction \in \inneractions \mid \safeconstr(\state, \outeraction, \inneraction) \geq \zeros\}$ via a vector-valued \mydef{safety constraint function} $\avoidconstr: \states^2 \times \states \times \states \to \R^\numconstrs$, which outputs a subset of the protagonist's actions in the safe set, i.e., for all $(\state, \outeraction) \in \states^2 \times \outeractions$, $\coupledactions(\state, \outeraction) \subseteq \{\inneraction \in  \inneractions \mid \detertrans (\state, \outeraction, \inneraction) \in \safestates\}$. 
Note that we do not require this inclusion to hold with equality; in this way, the protagonist can choose to restrict itself to actions far from the boundaries of the avoid set, thereby increasing its safety, albeit perhaps at the cost of liveness.
Overloading notation, the feasible policy correspondence $\coupledactions (\policy[\outeraction]) \doteq \{ \policy[\inneraction] : \states \to \inneractions \mid \policy[\inneraction] (\state) \in \coupledactions (\state, \policy[\outeraction] (\state)), \text{for all } \state \in \states \}$. 

We consider two forms of reward functions. 
The first, called the \mydef{reach probability reward}, $\reward (\state, \outeraction, \inneraction)  = \setindic[{\targetstates}] (\state[][\inneraction])$, is an indicator function that awards the protagonist with a payoff of $1$ if it enters the target set, and $0$ otherwise.
Under this reward function, the cumulative payoff function (i.e., the expected value of these rewards in the long term)
represents the probability that the protagonist reaches the target, hence its name. 
The second reward function is the \mydef{reach distance reward function}, $\reward (\state, \outeraction,  \inneraction)  = - \min_{\state[][][\prime] \in \targetstates} \, \| \state[][\inneraction] -\state[][][\prime] \|^2$, which penalizes the protagonist based on how far away it is from the target set.
For any policy profile $\policy \doteq (\policy[\outeraction], \policy[\inneraction]): \states^2 \to \outeractions \times \inneractions$ and choice of reward function $\reward$, the \mydef{cumulative payoff function} is given by $\cumulutil (\policy) \doteq \Ex_{\staterv[0] \sim \initstates} [\sum_{\iter = 0}^\infty \discount^\iter \reward (\staterv[\iter], \policy(\staterv[\iter])) \mid \staterv[\iter+1] = \detertrans (\staterv[\iter], \policy(\staterv[\iter]))]$. In what follows, we model the leader's policy $\policy[\outeraction] \doteq \policy[\outer]$ as parameterized by $\outer \in \outerset \subset \R^{\numstates \times \numstates}$, and the follower's policy $\policy[\inneraction] \doteq \policy[\inner]$ as parameterized by $\inner \in \innerset \subset \R^{\numstates \times \numstates}$. \samy{}{(Recall that $\numstates$ is the dimension of the state space.)}
\amy{this expression is a bit wonky. i don't get why you condition on $\staterv[\iter+1] = \detertrans (\staterv[\iter], \policy(\staterv[\iter]))$. i think you might instead want it as a subscript under the expectation.
here's what we have back in the intro:
\begin{align}
\Ex \left[\sum_{\iter = 0}^\infty \discount^\iter \reward(\staterv[\iter], \policy[\outeraction] (\staterv[\iter]), \policy[\inneraction] (\staterv[\iter])) \right]
\enspace ,
\end{align}
\noindent
where the expectation is with respect to $\staterv[0] \sim \initstates$ and $\staterv[\iter+1] \sim \trans (\cdot \mid \state[\iter], \policy[\outeraction] (\staterv[\iter]), \policy[\inneraction] (\staterv[\iter]))$.
}

\samy{A typical solution concept for reach-avoid games is \mydef{generalized Nash equilibrium} \cite{laine2023computation, margellos2011hamilton}\samy{: a policy profile $(\policy[\outeraction][][*], \policy[\inneraction][][*])$
s.t.\@ $\max_{\policy[\inneraction] \in \coupledactions (\policy[\outeraction][][*])}\cumulutil(\policy[\outeraction][][*], \policy[\inneraction]) \leq \cumulutil(\policy[\outeraction][][*], \policy[\inneraction][][*]) \leq \min_{\policy[\outeraction] \in \outeractions^\states} \cumulutil(\policy[\outeraction], \policy[\inneraction][][*])$}{}. 
Instead, we focus on (recursive) Stackelberg equilibrium (see \Cref{sec:policy_grad}).}{}

\amy{do we need to specify what happens if there are no feasible actions? namely that the protagonist has been captured, and earns 0 reward ever after? actually, is this what happens? or does the antagonist win, meaning the reward is then $-1$?} \deni{I don't think we need to specify, a solution just does not exist for such a game. Existence of Stackelberg requires you to always have a feasible action, which under this model is assurable.} \amy{but my guess is that the protagonist can get cornered in Arjun's experiments? so there, we may need to say something about this.}

\sdeni{
\begin{assumption}[Parameterization]
\label{assum:parameterization}
    1.~The players' parameter spaces $\outerset$ and
    $\innerset$ are non-empty, compact, and convex; and
    2.~the players use decentralized bilinear \amy{what does bilinear mean here? they both just look linear to me.} policies, i.e., $\policy[\outer] (\state) \doteq \outer \state[][\outeraction]$ and $\policy[\inner] (\state) \doteq \outer \state[][\inneraction]$.
    \deni{Moved this, convex-concave and parameterization goes hand in hand you cannot seperate them!!!}
\end{assumption}

Fixing the antagonist's policy, the protagonist's reach problem reduces to an MDP in which its payoffs depend only on its location in the state space; as such, a decentralized policy is optimal.
\samy{While this might not be the case for the antagonist in general, we note that our results hold when the policy of the antagonist depends on both and its and the protagonist's state.}{} 
Furthermore, when the target set is a singleton, the protagonist's best-response problem reduces to a linear quadratic regulator problem,\amy{add ref} for which this parameterization is again optimal.
}{}

The next assumption ensures that 
1.~under the reach probability reward function, a reach-avoid game is a convex-non-concave zero-sum Markov Stackelberg game (i.e., the marginal function $\outer \mapsto \marginalfunc (\outer)$ is convex, and the cumulative payoff function $\inner \mapsto \cumulutil (\outer, \inner)$ is non-concave, for all $\outer \in \outerset$), and 
2.~under the reach distance reward function, a reach-avoid game is a convex-concave zero-sum Markov Stackelberg game:
\deni{Maybe add something how these also guarantee the convex-concavity of the generalized Nash equilibrium problem.}

\begin{assumption}[Convex-Concave Reach-Avoid Game]
\label{assum:reach_avoid}
    1.~The state space $\states$ and the target set $\targetstates$ are non-empty and convex;
    2.~the displacement functions $\detertrans[\outeraction], \detertrans[\inneraction]$ are affine; and 
    3.~for all states $\state \in \states$, the safety set constraints $(\outeraction, \inneraction) \mapsto \safeconstr(\state, \outeraction, \inneraction)$ are log-convex-concave, i.e. $\safeconstr(\state, \outeraction, \inneraction)$ is log-convex in $\outeraction$, for all $\inneraction \in \inneractions$, and concave in $\inneraction$ for all $\outeraction \in \outeractions$.
    \sdeni{}{4.~The players' parameter spaces $\outerset$ and
    $\innerset$ are non-empty, compact, and convex; and
    5.~the players use decentralized bilinear  policies, i.e., $\policy[\outer] (\state) \doteq \outer \state[][\outeraction]$ and $\policy[\inner] (\state) \doteq \outer \state[][\inneraction]$.}
\end{assumption}

Part 1 is a standard assumption commonly imposed on reach-avoid games (see, for instance \citet{fisac2015reach}). 
Part 2 is satisfied by natural displacement functions of the type $\detertrans (\state, \outeraction, \inneraction) = \state + \beta(\outeraction, \inneraction)$, for some $\beta \in \R$, which is a natural characterization of all displacement functions with constant velocity $\beta$, when $\outeractions = \inneractions \subseteq \{ \z \in \states \mid \|\z \| \, = 1\}$. 
Part 3 is satisfied by various action correspondences, such as \deni{check that I appropriately subscript $\detertrans$ everywhere.} $\safeconstr(\state, \action, \inneraction) \doteq \exp\{ \min_{\state[][][\prime] \in \avoidstates} \, \| (\detertrans[\outeraction] (\state[][\outeraction], \outeraction), \state[][\inneraction]) - \state[][][\prime] \|\} - 1 - \| \detertrans[\inneraction] (\state[][\inneraction], \inneraction) - \state[][\inneraction]\|$, which shrinks the space of actions exponentially as the protagonist approaches the antagonist, and can thus be interpreted as describing a safety-conscious protagonist.
\amy{maybe add a reference, if you didn't invent this $\safeconstr$?}

\begin{restatable}{theorem}{thmreachavoid}
\label{thm:reach_avoid}
    Under the reach distance (resp.\@ reach probability) reward function, any reach-avoid game for which \Cref{assum:reach_avoid} hold is convex-concave (resp.\@ convex-non-concave). 
    Moreover, nested SGDA (resp.\@ saddle-point-oracle gradient descent) is guaranteed to converge in such games to recursive Stackelberg equilibrium policies in polynomial time.
\end{restatable}

\textbf{Experiments.}
We ran a series of experiments \sdeni{}{in which we initialized and trained neural policies in a reach-avoid game in a 2-dimensional state space with the antagonist and the protagonist modelled as Dubbins cars subject to a displacement function which is not affine  \cite{isaacs1954differential}.}%
\footnote{The details of our experimental setup can be found in the full version of the paper. Our code repository can be found at: \publiccoderepo.}
whose goal was to assess the performance of nested policy GDA with more complex, i.e., neural, policy parameterizations, \sdeni{}{and with non affine transitions} as well as the performance of \sdeni{Stackelberg equilibrium}{SE} as compared to \sdeni{Nash equilibrium}{GNE}. The target set is a select subset of the state space, while the avoid set---which defines the safe set and the feasible action and policy correspondences---is a ball around the antagonist.
An example trajectory is depicted in \Cref{fig:trajectory}.

\sdeni{We consider a variant of the two-player differential game introduced in \citet{isaacs1954differential}, played by two Dubins cars.
A Dubins car is a simplified model of a vehicle with a constant forward speed $c$ and a constrained turning radius, $\omega$.
Popular in control theory, they are used to model ground robots, aircrafts, and underwater vehicles \cite{Lav06}.

The state space in our game is a subset of the two-dimensional Euclidean plane, with the state of each car is given by $\state \doteq (x_1, y_1, x_2, y_2)$, representing the $x$ and $y$ coordinates of the front and rear of each car, respectively.
The cars' action spaces are the scalars $\{ -1, 0, 1 \}$, describing, respectively, a left turn of $\omega$ degrees, no turn, and a right turn of $\omega$ degrees.
%
%
%
%
The displacement function $\detertrans$ is an affine transformation comprising a rotation and a translation:
\begin{align*}
\detertrans[{\w}] (\state[][{\w}], \w) = 
v \w
\begin{bmatrix}
\cos(\omega) & -\sin(\omega) \\
\sin(\omega) & \cos(\omega) \\
\cos(\omega) & -\sin(\omega) \\
\sin(\omega) & \cos(\omega) 
\end{bmatrix}
\begin{bmatrix}
x_1 \\
y_1 \\
x_2 \\
y_2
\end{bmatrix}
\end{align*}
}{}



%
\sdeni{We experiment with only the reach distance, not the reach probability, reward function.
In all safe states, 
the reward is actually a penalty, measuring the protagonist's distance to the target set, while a bonus is awarded upon reaching a target, at which point the game ends.
This reward function suffices for our Stackelberg game setup, which enforces the hard constraint that the protagonist cannot move into the avoid set.
In our \sdeni{Nash}{GNE} game setup, we achieve a similar effect by enhancing the aforementioned reward function with a large penalty whenever the protagonist touches the avoid set.
Furthermore, as in the case of reaching the target, touching the avoid set ends the game. }{}

\sdeni{Our experiments were run on a 7x7 square grid, with the target a semi-circle of radius 1 centered along the lower edge, and the avoid set a circle of radius $0.3$ around the antagonist.
We set the bonus for reaching the target at $200$, the penalty for entering the avoid set at $-200$, $\omega = 30^\circ$, and $c = 0.25$.
\samy{}{We also end the game after 100\amy{i made this up!!!} iterations if the protagonist has not yet reached the target set.}}{}

\begin{wrapfigure}{R}{0.5\textwidth}
    \vspace*{-0.8cm}
    \centering
    \includegraphics[width=\textwidth/2]{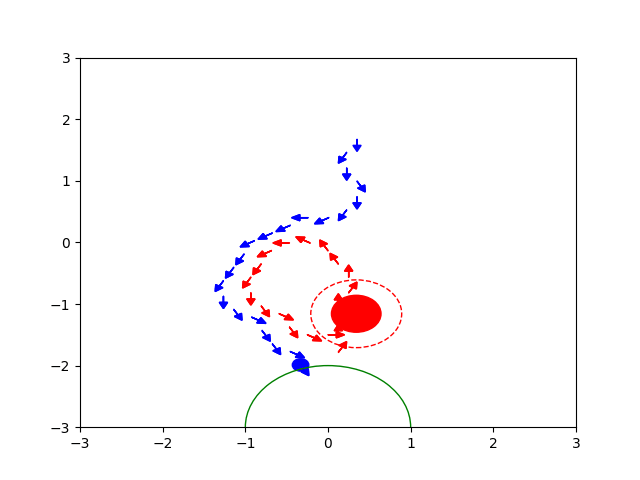}
    \caption{One run of the reach-avoid game. The protagonist (blue) attempts to reach the target set (green), while the antagonist (red) tries to prevent the protagonist doing so. The dotted line around the antagonist represents \samy{the region it can cover in one move}{the avoid set}. \amy{i mean, hopefully it represents the avoid set! who cares about the space it can cover in one move?}}
      \label{fig:trajectory}
    \vspace*{-0.5cm}
\end{wrapfigure}





Using this experimental setup, we train agents to play three games.
The first (resp.\@ second) is the Stackelberg (resp.\@ \sdeni{Nash}{simultaneous-move}) variant of the reach-avoid game, for which the training algorithm is stochastic policy GDA (resp.\@ SGDA).
In the third, the agents play a pursuit-evasion game,\amy{add ref} where the antagonist seeks to minimize its distance from the protagonist, and vice versa.
These agents play a Stackelberg game, and are trained using nested policy GDA.
We then evaluate the policies learned by each of the protagonists against those of the antagonists.

\Cref{tab:wins} shows the outcome of 50 match-ups between each pair of policies.
We find that the RA-Stackelberg protagonist wins (i.e., reaches the target set) more often than the \sdeni{Nash}{GNE} protagonist in all cases.
Furthermore, \Cref{tab:lengths} shows that the RA-Stackelberg protagonist is safer, since the game lasts longer before it is captured; and it is also livelier, since it wins more quickly than its \sdeni{Nash}{GNE} counterpart.

\amy{why isn't PE-Stackelberg (pursuit evasion) in \Cref{tab:lengths}? b/c it never wins!!!}

\amy{we need much more explanation of these numbers? why does Stackelberg beat Reverse every single time? why is it no contest? and why does Stackelberg also beat itself nearly every time? please add explanations of the numbers in the table to the text!}

\begin{table}[H]
\begin{minipage}{0.5\linewidth}
\resizebox{1\columnwidth}{!}{%
\begin{tabular}{|l|l|l|l|l|}
\hline
& \multicolumn{2}{c|}{Win Length} & \multicolumn{2}{c|}{Loss Length} \\
\hline
\textbf{Protagonist Type} & \textbf{Average} & \textbf{Std. Dev.} & \textbf{Average} & \textbf{Std. Dev.} \\ \hline
RA-Stackelberg & \textbf{28.46} & 1.84 & \textbf{32.56} & 7.52 \\ \hline
\sdeni{Nash}{GNE} & 32.45 & 2.38 & 18.72 & 5.65 \\ \hline
\end{tabular}
}
\subcaption{Average Game Lengths. \amy{this table makes no sense to me, since in our game formulation, the protagonist/attacker cannot lose!!! maybe there is a time-out eventually, but that is not described as part of our setup.}}
\label{tab:lengths}
\end{minipage}
\begin{minipage}{0.5\linewidth}
\resizebox{1\columnwidth}{!}{%
\begin{tabular}{|l|l|l|}
\hline
\textbf{Protagonist vs.\@ Antagonist} & \textbf{Avg.} & \textbf{Std. Dev.} \\ \hline
RA-Stackelberg vs.\@ \sdeni{Nash}{GNE} & \textbf{42.2} & 1.64 \\ 
\sdeni{Nash}{GNE} vs.\@ \sdeni{Nash}{GNE} & 30.4 & 2.41 \\ \hline 
RA-Stackelberg vs.\@ PE-Stackelberg & \textbf{50} & 0 \\ 
\sdeni{Nash}{GNE} vs.\@ PE-Stackelberg & 0 & 0 \\ \hline 
RA-Stackelberg vs.\@ RA-Stackelberg & \textbf{49.2} & 0.84 \\ 
\sdeni{Nash}{GNE} vs.\@ RA-Stackelberg & 9.6 & 3.78 \\ \hline
\end{tabular}
}
\subcaption{Average Number of Protagonist Wins}
\label{tab:wins}
\end{minipage}
\caption{Number of wins/losses for the protagonist and the associated game lengths.}
\end{table}

\section{Acknowledgments}
Denizalp Goktas was supported by a JP Morgan AI fellowship. 
Arjun Prakash was partially supported by ONR N00014-22-1-2592 and the Quad Fellowship.

\bibliographystyle{plainnat} 
\bibliography{references}  

\newpage
\appendix 
should be submitted separately
\section{Preliminaries}\label{sec_app:prelims}

We use caligraphic uppercase letters to denote sets (e.g., $\calX$), 
bold lowercase letters to denote vectors (e.g., $\price$), lowercase letters to denote scalar quantities (e.g., $x$), and uppercase letters to denote random variables (e.g., $X$).
We denote functions by a letter determined by the value of the function, e.g., $f$ if the mapping is scalar valued, $\f$ if the mapping is vector valued, and $\calF$ if the mapping is set-valued.
We denote the set of integers $\left\{1, \hdots, n\right\}$ by $[n]$, the set of natural numbers by $\N$, and the set of real numbers by $\R$. 
We denote the positive and strictly positive  (resp. negative and strictly negative) elements of a set by $+$ and $++$ subscripts, respectively, e.g., $\R_+$ and $\R_{++}$ (resp. $\R_-$ and $\R_{--}$). 
\section{Related Work}
\label{sec:related}
\paragraph{Related Work}

The computation of Stackelberg equilibrium in two-player stochastic Stackelberg games has been studied in various interesting settings.
\citeauthor{bensoussan2015maximum} (\citeyear{bensoussan2015maximum}) study continuous-time general-sum stochastic Stackelberg games with continuous action spaces, and prove the existence of a solution \amy{a solution, generically? not an eqm?} in this setting.
\citeauthor{vorobeychik2012computing} (\citeyear{vorobeychik2012computing}) consider a general-sum
stochastic Stackelberg game with finite state-action spaces \amy{finite states as well?} and an infinite horizon.
They show that stationary Stackelberg equilibrium policies do not exist in this very general setting, but nonetheless identify a subclass of games, namely team (or potential) Stackelberg games for which stationary Stackelberg equilibrium policies do exist.
\citeauthor{vu2022stackelberg} (\citeyear{vu2022stackelberg}) study the empirical convergence of policy gradient methods in the same setting \amy{also in team games, i presume?} as \citeauthor{vorobeychik2012computing}, while \citeauthor{ramponi2022learning} \cite{ramponi2022learning} study non-stationary equilibria in this setting \amy{team games, or more general?}, assuming a finite horizon.
\citeauthor{chang2015leader} (\citeyear{chang2015leader}) and \citeauthor{sengupta2020multi} (\citeyear{sengupta2020multi}) consider a partially observable version of \citeauthor{vorobeychik2012computing}'s \samy{}{team game} \amy{???} model, and provide methods to compute Stackelberg equilibria in their setting. \citeauthor{bai2021sample} (\citeyear{bai2021sample}) consider a related model, the Bandit-RL model, in which the leader chooses an action, which defines a discrete state-action MDP for the follower.


In none of the aforementioned models is the follower's action constrained by the leader's choice.
\citet{goktas2021minmax} were the first to study min-max games with such dependencies, proposing two polynomial-time algorithms, nested gradient descent ascent (GDA) \cite{goktas2021minmax} and simultaneous GDA for such convex-concave games \cite{goktas2021robustminmax}.
\citet{goktas2022zero} extended these ideas to convex-concave zero-sum stochastic games, solving the planning problem (assuming knowledge of rewards and transitions) via value iteration, whereas we solve the learning problem via policy gradient, learning only from observed trajectories of play, i.e., noisy gradients.

Some recent research concerns Stackelberg games with many followers.
\citeauthor{vasal2020stochastic} (\citeyear{vasal2020stochastic}) presents algorithms to solve discrete-time, finite horizon one-leader many-followers stochastic Stackelberg games
with discrete action and state spaces.
\citeauthor{demiguel2009stochastic} (\citeyear{demiguel2009stochastic}) study a stochastic Stackelberg game-like market model with $n$ leaders and $m$ followers; they prove the existence of a Stackelberg equilibrium in their model, and devise algorithms (without theoretical guarantees) that converge to such an equilibrium in their experiments.

Stochastic 
Stackelberg games \cite{li2017review} have been used to model a wide range of problems, including security \cite{vasal2020stochastic, vorobeychik2012computing}, insurance provision \cite{chen2018new, yuan2021robust}, advertising \cite{he2008cooperative}, robust agent design \cite{rismiller2020stochastic}, and resource allocation across time and intertemporal pricing \cite{oksendal2013stochastic}. 

\newpage
\section{Ommited Algorithm Details}
\label{sec:algos}


\begin{algorithm}[H]
\textbf{Inputs:} $\outerset, \innerset, \util, \outer[][][][0], \inner[][][][0], \langmult[][][0], \numiters[\outer], \numiters[\inner], \left\{\learnrate[\outer][\iter] \right\}_{\iter = 0}^{\numiters[\outer]-1}, \left\{\learnrate[\inner][\iter] \right\}_{\iter = 0}^{\numiters[\outer]-1}$\\
\textbf{Outputs:} $(\outer[][][][\iter], \inner[][][][\iter], \langmult[][][\iter])_{\iter = 0}^{\numiters[\outer]-1}$ 
\begin{algorithmic}[1]
\caption{Nested Policy Gradient Descent Ascent 
\label{alg:nested_reinforce}}
\State $\inner[][][][] \gets  \inner[][][][0]$
\State $\langmult[][][] \gets  \langmult[][][0]$
\For{$\iter = 0, \hdots, \numiters[\outer] - 1$}
        \For{$s = 0, \hdots, \numiters[\inner] - 1$}
            \State Sample trajectory $\hist[][][]  \sim \histdistrib[][{(\outer[][][][\iter], \inner[][][][\iter])}]$ and observe rewards $\rewards = (\rewards[][\numhorizon])_\numhorizon$ \amy{why the subscript?}
            \State $\inner[][][] \gets  \project[\innerset] \left[\inner[][][] + \grad[\inner] \lang[{\hist}](\inner[][][], \langmult[][]; \outer[][][][\iter]) \right]$
            \State  $\langmult[][][] \gets  \project[\langmults]\left[\langmult[][][] - \grad[\langmult] \lang[{\hist}](\inner[][][][], \langmult[][][]; \outer[][][][\iter]) \right]$
        \EndFor
    \State $\inner[][][][\iter] \gets  \inner[][][][]$
    \State $\langmult[][][\iter] \gets  \langmult[][][]$
    \State Sample trajectory $\hist[][]  \sim \histdistrib[][{(\outer[][][][\iter], \inner[][][][\iter])}]$ and observe rewards $\rewards = (\rewards[][\numhorizon])_{\numhorizon}$ \amy{why the subscript?}
    \State $\outer[][][][\iter + 1] \gets \project[\outerset]\left[\outer[][][][\iter] +  \grad[\outer] \lang[{\hist}](\inner[][][][\iter], \langmult[][][\iter]; \outer[][][][\iter]) \right]$
\EndFor
\State \Return $(\outer[][][][\iter], \inner[][][][\iter], \langmult[][][\iter])_{\iter = 0}^{\numiters[\outer]-1}$ 
\end{algorithmic}
\end{algorithm}


\begin{algorithm}[H]
\textbf{Inputs:} $\outerset, \innerset, \util, \outer[][][][0], \inner[][][][0], \numiters[\outer], \numiters[\inner], \left\{\learnrate[\outer][\iter] \right\}_{\iter = 0}^{\numiters-1}, \left\{\learnrate[\inner][\iter] \right\}_{\iter = 0}^{\numiters-1}, \hat{v}, \boldsymbol{w}, \left\{\learnrate[\boldsymbol{w}][\iter] \right\}_{\iter = 0}^{\numiters-1} $\\
\textbf{Outputs:} $(\outer[][][][\iter], \inner[][][][\iter], \langmult[][][\iter])_{\iter = 0}^{\numiters-1}$ 
\begin{algorithmic}[1]
\caption{Nested REINFORCE with baselines 
\label{alg:sim_reinforce}}
\State $\inner[][][][] \gets  \inner[][][][0]$
\State $\langmult[][][] \gets  \langmult[][][0]$
\For{$\iter = 0, \hdots, \numiters[\outer] - 1$}
        \For{$s = 0, \hdots, \numiters[\inner] - 1$}
        \amy{why are there two loops in this algo?}
            \State Sample trajectory $\hist[][][]  \sim \histdistrib[][{(\outer[][][][\iter], \inner[][][][\iter])}]$ and observe rewards $\rewards = (\rewards[][\numhorizon])_\numhorizon$ \amy{why the subscript?}
            \For{Each step of the episode $t = t_0,1...,$}
                \State $G \gets \sum^T_{k=t+1} \rewards[][\numhorizon]$
                \State $\delta \gets G - \hat{v}(\state[t], \boldsymbol{w})$
                \State $\inner[][][][\iter] \gets  \inner[][][] - \learnrate[\inner] \gamma \delta \grad[\inner] \ln \policy[\inner](A_t ; \state[t],\inner)$

        \EndFor
    \EndFor

    \For{Each step of the episode $t = t_0,1...$}
    \State Sample trajectory $\hist[][][]  \sim \histdistrib[][{(\outer[][][][\iter], \inner[][][][\iter])}]$ and observe rewards $\rewards = (\rewards[][\numhorizon])_\numhorizon$
    \State $G \gets \sum^T_{k=t+1} \rewards[][\numhorizon]$
    \State $\delta \gets G - \hat{v}(\state[t], \boldsymbol{w})$
    \State $\outer[][][][\iter] \gets  \outer[][][] + \learnrate[\outer] \gamma \delta \grad[\outer] \ln \policy[\outer](A_t ; \state[t],\outer)$
    \State $\boldsymbol{w}^t \gets \boldsymbol{w} + \learnrate[\boldsymbol{w}] \gamma \grad \hat{v}(\state[t], \boldsymbol{w})$ 
    \EndFor
\EndFor

\State \Return $(\outer[][][][\iter], \inner[][][][\iter])_{\iter = 0}^{\numiters-1}$ 
\end{algorithmic}
\end{algorithm}


\begin{algorithm}[H]
\textbf{Inputs:} $\outerset, \innerset, \util, \outer[][][][0], \inner[][][][0], \langmult[][][0], \numiters, \left\{\learnrate[\outer][\iter] \right\}_{\iter = 0}^{\numiters-1}, \left\{\learnrate[\inner][\iter] \right\}_{\iter = 0}^{\numiters-1}$\\
\textbf{Outputs:} $(\outer[][][][\iter], \inner[][][][\iter], \langmult[][][\iter])_{\iter = 0}^{\numiters-1}$ 
\begin{algorithmic}[1]
\caption{Simultaneous Policy Gradient Descent Ascent 
\label{alg:sim_pgda}}
\State $\inner[][][][] \gets  \inner[][][][0]$
\State $\langmult[][][] \gets  \langmult[][][0]$
\For{$\iter = 0, \hdots, \numiters - 1$}
        \amy{why are there two loops in this algo?}
        \State Sample trajectory $\hist[][][]  \sim \histdistrib[][{(\outer[][][][\iter], \inner[][][][\iter])}]$ and observe rewards $\rewards = (\rewards[][\numhorizon])_\numhorizon$ \amy{why the subscript?}
        \State $\inner[][][] \gets  \project[\innerset] \left[\inner[][][] + \grad[\inner] \lang[{\hist}](\inner[][][], \langmult[][]; \outer[][][][\iter]) \right]$
        \State  $\langmult[][][] \gets  \project[\langmults]\left[\langmult[][][] - \grad[\langmult] \lang[{\hist}](\inner[][][][], \langmult[][][]; \outer[][][][\iter]) \right]$
        \State $\inner[][][][\iter] \gets  \inner[][][][]$
        \State $\langmult[][][\iter] \gets  \langmult[][][]$
        \State $\outer[][][][\iter + 1] \gets \project[\outerset]\left[\outer[][][][\iter] +  \grad[\outer] \lang[{\hist}](\inner[][][][\iter], \langmult[][][\iter]; \outer[][][][\iter]) \right]$
        \EndFor
\State \Return $(\outer[][][][\iter], \inner[][][][\iter], \langmult[][][\iter])_{\iter = 0}^{\numiters-1}$ 
\end{algorithmic}
\end{algorithm}


\begin{algorithm}[H]
\textbf{Inputs:} $\outerset, \innerset, \util, \outer[][][][0], \inner[][][][0], \numiters, \left\{\learnrate[\outer][\iter] \right\}_{\iter = 0}^{\numiters-1}, \left\{\learnrate[\inner][\iter] \right\}_{\iter = 0}^{\numiters-1}, \hat{v}, \boldsymbol{w}, \left\{\learnrate[\boldsymbol{w}][\iter] \right\}_{\iter = 0}^{\numiters-1} $\\
\textbf{Outputs:} $(\outer[][][][\iter], \inner[][][][\iter], \langmult[][][\iter])_{\iter = 0}^{\numiters-1}$ 
\begin{algorithmic}[1]
\caption{Simultaneous REINFORCE with baselines 
\label{alg:sim_reinforce}}
\State $\inner[][][][] \gets  \inner[][][][0]$
\State $\langmult[][][] \gets  \langmult[][][0]$
\For{$\iter = 0, \hdots, \numiters - 1$}
        \amy{why are there two loops in this algo?}
        \State Sample trajectory $\hist[][][]  \sim \histdistrib[][{(\outer[][][][\iter], \inner[][][][\iter])}]$ and observe rewards $\rewards = (\rewards[][\numhorizon])_\numhorizon$ \amy{why the subscript?}
        \For{Each step of the episode $t = t_0,1...,$}
        \State $G \gets \sum^T_{k=t+1} \rewards[][\numhorizon]$
        \State $\delta \gets G - \hat{v}(\state[t], \boldsymbol{w})$
        \State $\boldsymbol{w}^t \gets \boldsymbol{w} + \learnrate \gamma \grad \hat{v}(\state[t], \boldsymbol{w})$ 
        \State $\inner[][][][\iter] \gets  \inner[][][] - \learnrate[\inner] \gamma \delta \grad[\inner] \ln \policy[\inner](A_t ; \state[t],\inner)$
        \State $\outer[][][][\iter] \gets  \outer[][][] + \learnrate[\outer] \gamma \delta \grad[\outer] \ln \policy[\outer](A_t ; \state[t],\outer)$

        \EndFor
        
        \EndFor
\State \Return $(\outer[][][][\iter], \inner[][][][\iter])_{\iter = 0}^{\numiters-1}$ 
\end{algorithmic}
\end{algorithm}

To invoke the deterministic policy gradient theorem requires access to two gradients.
The gradient $\grad[\outer] \policy[\outer] (\staterv)$ of the policy w.r.t.\@ its parameters is available by design in all differentiable parameterizations of interest. 
The gradient $\grad[(\outeraction, \inneraction)] \actionvalue[\outer, \inner]$ of the state-action value function w.r.t\@ both players' actions can be computed via the chain rule, since our oracle affords us access to the gradients of both the reward and probability transition functions: 
$\grad[(\outeraction, \inneraction)] \actionvalue[\outer, \inner](\state, \outeraction, \inneraction) = \Ex_{\hist} \left[\gradactionvalue[](\state, \outeraction, \inneraction, \outer, \inner; \hist) \right]$\sdeni{, where 
$\gradactionvalue[](\state, \outeraction, \inneraction, \outer, \inner; \hist) \doteq \samy{(\gradactionvalue[\outer] \gradactionvalue[\inner]) (\state, \outeraction, \inneraction, \outer, \inner; \hist) \doteq}{}$ \amy{i don't think we need this second bit. at least not in the main paper.} \deni{Ok will move this bit when I get to appendix tmrw.}
\begin{align*}
&\grad[(\outeraction, \inneraction)] \reward(\state[0], \outeraction, \inneraction) + \discount \grad[(\outeraction, \inneraction)] \trans(\state[1][] \mid \state[0][][], \outeraction, \inneraction)
\left[ \reward\left(\state[1][][], \policy[\outer](\state[1]), \policy[\inner] (\state[1]) \right) + \right. \\
&\left. \sum_{\iter = 2}^\infty \reward\left(\state[\iter], \policy[\outer](\state[\iter]), \policy[\inner] (\state[\iter]) \right) \prod_{k = 2}^{\iter} \discount^{k+1} \trans(\state[k][] \mid \state[k-1], \policy[\outer](\state[k-1]), \policy[\inner] (\state[k-1])) \right] 
\enspace .
\end{align*}
}{.}
\newpage
\section{Omitted Proofs}
\label{sec_app:proofs}

\begin{lemma}[Convex-Concave Assumption]\amy{why is this called an Assumption?}
\label{lemma:convex_concave_proof}
    Suppose that the min-max Stackelberg game $(\outerset, \innerset, \obj, \constr)$ satisfies \Cref{assum:convex_concave}, then the game is convex-concave. 
\end{lemma}

\begin{proof}
    By Proposition 3.1 of \citeauthor{fiacco1986convexity} \cite{fiacco1986convexity}, under \Cref{assum:convex_concave}, $\val$ is convex in $\outer$. In addition, since by assumption $\obj$ is concave in $\inner$, the game is convex-concave.
\end{proof}

\begin{lemma}[Alternative Convex-Concave Assumption]
    
\end{lemma}

\subsection{Algorithm for Convex-Concave Min-Max Stackelberg Games}

\textbf{\sdeni{Proof overview}{Technical details}.}
\deni{If we are out of space all of this below can imo move to appendix.}

\sdeni{Intuitively, saddle-point-oracle SGD and nested SGDA should compute a $(\varepsilon, \delta)$-Stackelberg equilibrium; however, there are three technical caveats, the first two of which concern the analysis of saddle-point-oracle SGD (and hence by consequence nested SGDA  which may prevent this intuitive approach from succeeding.
We explain and resolve them each in turn.}

\sdeni{}{
First, while we know that in convex-concave games the marginal function $\val$ is convex, to guarantee the convergence of saddle-point-oracle SGD even with an exact oracle, at the least, we need $\val$ to be Lipschitz-continuous (i.e., have bounded subgradients). Second, as we do not necessarily have access to an exact saddle-point oracle, we can only approximate a subgradient of $\val$ at a given $\outerpoint \in \outerset$ with $\widehat{\grad \val}$, and hence we have to ensure that approximation errors do not accumulate during the course of running the algorithm. Finally, when we implement the oracle of saddle-point-oracle SGD using SGDA, to guarantee the worst case convergence of SGDA to a saddle point, we need to define a non-empty, compact, and convex set $\langmults^*(\outer) \subseteq \langmults \subseteq \R^\numconstrs_+$ containing the optimal Lagrange multiplier associated with any $\outer \in \outerset$ to run the SGDA with so that the Lagrange multipliers computed by the nested SGDA remain always bounded.}

\deni{Bounding the approximation of the subgradient error (submetric regularity) 
Lipschitz Coefficient of $\val$.

Set of Langran multiplier constraints.  .}

\sdeni{}{We first resolve the third issue as it also implies a solution for the first. To this end we extend \citeauthor{nedic2009approximate} \cite{nedic2009approximate}'s result that provides for any convex optimization problem satisfying Slater's condition a choice of non-empty, compact, and convex set which is guaranteed to contain all optimal Lagrange multipliers to a class of convex optimization. In particular, we resolve the third issue by setting $\langmults = \{ \langmult \in \R_+^\numconstrs \mid \forall \outer \in \outerset, \inner \in \argmax_{\inner[][][\prime] \in \innerset} \min_{\numconstr} \constr[][\numconstr](\outer, \inner[][][\prime]), \left\|\langmult \right\|_1 \leq \frac{ \obj(\outer, \innerpoint) - \max_{\innerpoint \in \innerset}\lang (\innerpoint, \langmult[][]; \outer)}{\min_{\numconstr} \constr[ ][\numconstr] (\outer, \inner)}\}$, which is a non-empty, compact, convex set which which can be shown to contain the optimal Lagrange multiplier associated with any leader action using Lemma 3 of \citet{nedic2009gda}. From this result, we directly obtain that $\val$ is $\lipschitz[\val]$-Lipschitz where $\lipschitz[\val] \leq \lipschitz[\lang] \doteq 
\max_{(\inner, \langmult, \outer) \in \innerset \times \langmults \times \outerset } \|\grad[\outer] \lang (\inner, \langmult; \outer)\|$.

Regarding, the second issue, that when the objective $\obj$ and constraints $\constr$ are Lipschitz-smooth, then the gradient approximation error at each iteration of saddle-point-oracle SGD can be shown to be bounded by the distance between the $\delta$-saddle point oracle and the set of optimal saddle points. In turn, we can show that Slater's condition, implies the submetric regularity constraint qualification (see, for instance \citet{mohammadi2022penalty} for background, and Proposition 3.1. of \citet{mohammadi2022variational} for a proof) which in turn allows us to to bound this distance as a function of $\delta$.}

The following lemma tells us that the approximation error in the gradient of the value function depends on the distance to the optimal Stackelberg equilibrium strategy for the leader as well as the optimal KKT point for the Lagrangian.
In particular, as the distance to the leader's Stackelberg equilibrium strategy decreases, the error in the gradient decreases, even if the distance to the optimal KKT multiplier is large.

\begin{lemma}[Gradient Approximation Error]
\label{lemma:error_bound}
    Consider a convex-concave min-max Stackelberg game $(\outerset, \innerset, \obj, \constr)$.
    Suppose that \Cref{assum:smooth} holds and that \Cref{alg:nested_sgda} is run on any inputs, and outputs $(\outer[][][][\iter], \inner[][][][\iter])_{\iter = 0}^{\numiters[\outer]-1}$. 
    For any $\iter \in [\numiters[\outer]]$, let $(\inner[][][*][\iter], \langmult[][*][\iter])$ be a saddle point of $\lang(\outer[][][][\iter], \inner[][], \langmult[])$, i.e., a solution to $\max_{\inner[][][] \in \innerset: \constr(\outer[][][][\iter], \inner[][][\prime]) \geq \zeros} \min_{\langmult[][] \in \R_+^\numconstrs} \lang(\outer[][][][\iter], \inner[][], \langmult[])$, and let $\outer[][][*][] \in \argmin_{\outer \in \outerset} \val(\outer)$. 
    Then:
        $\Ex\left[\left< \grad\val(\outer[][][][\iter]) - \grad[\outer] \lang[\obji, \constri](\outeraction[][][][\iter]), \outer[][][][\iter] - \outer[][][*][] \right> \right] \leq  \lipschitz[\lang] \left( \left\| \inner[][][*][\iter] -\inner[][][][\iter]\right\| + \left\| \langmult[][*][\iter] - \langmult[][][\iter] \right\|\right)  \left\|\outer[][][][\iter] - \outer[][][*][] \right\|$.
\end{lemma}

\begin{proof}[Proof of \Cref{lemma:error_bound}]
    \deni{Add the metric regularity bound here too as well.}Fix any $\iter \in [\numiters[\outer]]$. For conciseness, let $\outeraction[][][][\iter] \doteq (\outer[][][][\iter], \inner[][][][\iter], \langmult[][][\iter])$. Using the subdifferential envelope theorem \cite{goktas2021minmax}, and Cauchy-Schwarz, we have: 
\begin{align}
    \Ex\left[\left<\err[\iter], \outer[][][][\iter] - \outer[][][*][] \right> \right]&= \Ex\left[\left< \grad\val(\outer[][][][\iter]) - \grad[\outer] \lang[\obji, \constri](\outeraction[][][][\iter]), \outer[][][][\iter] - \outer[][][*][] \right> \right]\\
    &= \Ex\left[\left< \grad[\outer] \Ex[\lang[\obji, \constri](\outer[][][][\iter], \inner[][][*][\iter], \langmult[][*][\iter])] - \grad[\outer] \lang[\obji, \constri](\outeraction[][][][\iter]), \outer[][][][\iter] - \outer[][][*][] \right> \right]\\
    &= \Ex\left[\left< \grad[\outer] \lang[\obji, \constri](\outer[][][][\iter], \inner[][][*][\iter], \langmult[][*][\iter]) - \grad[\outer] \lang[\obji, \constri](\outeraction[][][][\iter]), \outer[][][][\iter] - \outer[][][*][] \right> \right]\\
    &\leq \Ex\left[\left\| \grad[\outer] \lang[\obji, \constri](\outer[][][][\iter], \inner[][][*][\iter], \langmult[][*][\iter]) - \grad[\outer] \lang[\obji, \constri](\outeraction[][][][\iter]) \right\| \left\|\outer[][][][\iter] - \outer[][][*][] \right\| \right]
\end{align}
Under \Cref{assum:smooth} $\lang[\obji, \constri]$ is Lipschitz-smooth since the objective $\obj[\obji]$ and constraint $\constr[\constri]$ functions are Lipschitz-smooth.
\begin{align}
    &\leq \Ex\left[\left\| \grad \lang[\obji, \constri](\outer[][][][\iter], \inner[][][*][\iter], \langmult[][*][\iter]) - \grad \lang[\obji, \constri](\outeraction[][][][\iter]) \right\| \left\|\outer[][][][\iter] - \outer[][][*][] \right\| \right]\\
    &\leq \Ex\left[\lipschitz[\lang] \left\| \left(\outer[][][][\iter], \inner[][][*][\iter], \langmult[][*][\iter] \right) - \left(\outer[][][][\iter], \inner[][][][\iter], \langmult[][][\iter] \right)\right\| \left\|\outer[][][][\iter] - \outer[][][*][] \right\| \right]\\
    &\leq \Ex\left[\lipschitz[\lang] \left( \left\| \inner[][][*][\iter] -\inner[][][][\iter]\right\| + \left\| \langmult[][*][\iter] - \langmult[][][\iter] \right\|\right)  \left\|\outer[][][][\iter] - \outer[][][*][] \right\| \right]\\
    &\leq \lipschitz[\lang] \left( \left\| \inner[][][*][\iter] -\inner[][][][\iter]\right\| + \left\| \langmult[][*][\iter] - \langmult[][][\iter] \right\|\right)  \left\|\outer[][][][\iter] - \outer[][][*][] \right\| 
\end{align}
where $(\inner[][][*][\iter], \langmult[][*][\iter])$ is a saddle point of $\lang(\outer[][][][\iter], \inner[][], \langmult[])$.
\end{proof}

\thmminmaxconvergence*

\begin{proof}[Proof of \Cref{thm:min_max_convergence}]
First, note that by Theorem 3.15 of \citeauthor{nemirovski2009robust} \cite{nemirovski2009robust}, the saddle point oracle can be achieved in $\tilde{O}(\nicefrac{1}{\delta^2})$ iterations, i.e., the inner loop of nested SGDA can compute a saddle point with $\delta$-saddle point residual in $\tilde{O}(\nicefrac{1}{\delta^2})$ iterations. The iteration complexities in the theorem can then be realized combining this iteration complexity with the results below.

Additionally, note that we have:
\begin{align}
\Ex_{\substack{\obji \sim \objdistrib\\ \constri \sim \constrdistrib}}\left[ \grad[\outer] \lang[\obji, \constri](\inner[][][][\iter], \langmult[][][\iter]; \outer[][][][\iter])\right] = \grad[\outer] \Ex_{\substack{\obji \sim \objdistrib\\ \constri \sim \constrdistrib}}\left[ \lang[\obji, \constri](\inner[][][][\iter], \langmult[][][\iter]; \outer[][][][\iter])\right] = \grad[\outer] \lang(\inner[][][][\iter], \langmult[][][\iter]; \outer[][][][\iter]) 
\end{align}


Conditioning on $\outer[][][][\iter]$, for any  $\outer[][][] \in \outerset$ and $\iter \in \iters$, we have in expectation:
\begin{align}
    &\Ex\left[ \left\| \outer[][][][\iter + 1] - \outer \right\|^2 \right] \\ 
    &=\Ex\left[ \left\| \project[\outerset] \left( \outer[][][][\iter] - \learnrate[\outer][\iter] \grad[\outer] \lang[{\obji, \constri}](\inner[][][][\iter], \langmult[][][\iter]; \outer[][][][\iter])\right) - \project[\outerset] \left( \outer \right) \right\|^2 \right]\\
    &\leq \Ex\left[\left\| \outer[][][][\iter] - \learnrate[\outer][\iter] \grad[\outer] \lang[{\obji, \constri}](\inner[][][][\iter], \langmult[][][\iter]; \outer[][][][\iter]) -  \outer \right\|^2 \right]\\
    &= \Ex\left[ \left\| \outer[][][][\iter] - \outer \right\|^2 - 2\learnrate[\outer][\iter] \left<\grad[\outer] \lang[{\obji, \constri}](\inner[][][][\iter], \langmult[][][\iter]; \outer[][][][\iter]),\left( \outer[][][][\iter] - \outer \right) \right> + (\learnrate[\outer][\iter])^2 \left\| \grad[\outer] \lang[{\obji, \constri}](\inner[][][][\iter], \langmult[][][\iter]; \outer[][][][\iter]) \right\|^2 \right]\enspace ,
\end{align}
where the first line follows from the stochastic subgradient method and the fact that $\outer \in \outerset$;
the second, because the project operator is a non-expansion; and
the third, by the definition of the norm.

From the linearity of the expectation, we then have:
{\small
\begin{align}
    &= \Ex\left[ \left\| \outer[][][][\iter] - \outer \right\|^2 \right] - 2\learnrate[\outer][\iter] \Ex\left[ \left< \grad[\outer] \lang[{\obji, \constri}](\inner[][][][\iter], \langmult[][][\iter]; \outer[][][][\iter]) ,  \outer[][][][\iter] - \outer \right>  \right]+ (\learnrate[\outer][\iter])^2 \Ex\left[ \left\| \grad[\outer] \lang[{\obji, \constri}](\inner[][][][\iter], \langmult[][][\iter]; \outer[][][][\iter]) \right\|^2 \right]\\
    &= \Ex\left[  \left\| \outer[][][][\iter] - \outer \right\|^2 \right] - 2\learnrate[\outer][\iter] \Ex\left[ \left< \grad[\outer] \lang(\inner[][][][\iter], \langmult[][][\iter]; \outer[][][][\iter]),  \outer[][][][\iter] - \outer \right>\right] + (\learnrate[\outer][\iter])^2 \Ex\left[ \left\| \grad[\outer] \lang[{\obji, \constri}](\inner[][][][\iter], \langmult[][][\iter]; \outer[][][][\iter]) \right\|^2 \right] \\
    &= \Ex\left[  \left\| \outer[][][][\iter] - \outer \right\|^2 \right] - 2\learnrate[\outer][\iter] \Ex\left[ \left< \grad[\outer] \lang(\inner[][][][\iter], \langmult[][][\iter]; \outer[][][][\iter]),  \outer[][][][\iter] - \outer \right>\right] + (\learnrate[\outer][\iter])^2 \Ex\left[ \left\| \grad[\outer] \obj[\obji] (\outer[][][][\iter], \inner[][][][\iter]) + \left< \langmult[][][\iter],  \grad[\outer] \constr[\constri] (\outer[][][][\iter], \inner[][][][\iter]) \right> \right\|^2 \right] \\
    &\leq \Ex\left[  \left\| \outer[][][][\iter] - \outer \right\|^2 \right] - 2\learnrate[\outer][\iter] \Ex\left[ \left< \grad[\outer] \lang(\inner[][][][\iter], \langmult[][][\iter]; \outer[][][][\iter]),  \outer[][][][\iter] - \outer \right>\right] + (\learnrate[\outer][\iter])^2 \Ex\left[ \left\| \grad[\outer] \obj[\obji] (\outer[][][][\iter], \inner[][][][\iter])\right\|^2 + \left\|\left< \langmult[][][\iter],  \grad[\outer] \constr[\constri] (\outer[][][][\iter], \inner[][][][\iter]) \right> \right\|^2 \right] \\
    &\leq \Ex\left[  \left\| \outer[][][][\iter] - \outer \right\|^2 \right] - 2\learnrate[\outer][\iter] \Ex\left[ \left< \grad[\outer] \lang(\inner[][][][\iter], \langmult[][][\iter]; \outer[][][][\iter]),  \outer[][][][\iter] - \outer \right>\right] + (\learnrate[\outer][\iter])^2 \Ex\left[ \left\| \grad[\outer] \obj[\obji] (\outer[][][][\iter], \inner[][][][\iter])\right\|^2 + \left\| \langmult[][][\iter] \right\|^2 \left\| \grad[\outer]\constr[\constri] (\outer[][][][\iter], \inner[][][][\iter]) \right\|^2 \right]\\
    &= \Ex\left[  \left\| \outer[][][][\iter] - \outer \right\|^2 \right] - 2\learnrate[\outer][\iter] \Ex\left[ \left<\grad[\outer] \lang(\inner[][][][\iter], \langmult[][][\iter]; \outer[][][][\iter]), \outer[][][][\iter] - \outer \right>\right] + (\learnrate[\outer][\iter])^2 \left( \variance[\grad \obj] + \langmultmax\variance[\grad \constr] \right)
    \enspace ,
 \end{align}
}
where $\langmultmax \doteq \max_{\langmult \in \langmults} \| \langmult\|^2$.

Define the gradient approximation error: $\err[\iter] \doteq \grad  \val(\outer[][][][\iter]) - \grad[\outer] \lang[\obji, \constri](\outeraction[][][][\iter])$, we then have: 
\begin{align}
    &=\Ex\left[  \left\| \outer[][][][\iter] - \outer \right\|^2 \right] - 2 \learnrate[\outer][\iter]  \Ex\left[ \left<\grad \val(\outer[][][][\iter]), \outer[][][][\iter] - \outer \right>\right]  + (\learnrate[\outer][\iter])^2 \left( \variance[\grad \obj] + \langmultmax\variance[\grad \constr] \right) + 2 \learnrate[\outer][\iter] \Ex\left[ \left<\err[\iter], \outer[][][][\iter] - \outer \right>\right]
\end{align}

Define $\xi^{(\iter)} \doteq \lipschitz[\lang] \left( \left\| \inner[][][*][\iter] -\inner[][][][\iter]\right\| + \left\| \langmult[][*][\iter] - \langmult[][][\iter] \right\|\right)  \left\|\outer[][][][\iter] - \outer[][][*][] \right\|$, by \Cref{lemma:error_bound}, we have:

{\small
\begin{align}
    &\leq \Ex\left[  \left\| \outer[][][][\iter] - \outer \right\|^2 \right] - 2 \learnrate[\outer][\iter]   \Ex\left[ \left<\grad \val(\outer[][][][\iter]), \outer[][][][\iter] - \outer \right>\right]  + (\learnrate[\outer][\iter])^2 \left( \variance[\grad \obj] + \langmultmax\variance[\grad \constr] \right)  \notag \\ 
    &+  2 \learnrate[\outer][\iter] \lipschitz[\lang] \left\|\outer[][][][\iter] - \outer[][][*][] \right\| \left( \left\| \inner[][][*][\iter] -\inner[][][][\iter]\right\| + \left\| \langmult[][*][\iter] - \langmult[][][\iter] \right\|\right)\\
    &\leq \Ex\left[  \left\| \outer[][][][\iter] - \outer \right\|^2 \right] - 2 \learnrate[\outer][\iter]  \Ex\left[ \left<\grad \val(\outer[][][][\iter]), \outer[][][][\iter] - \outer \right>\right]  + (\learnrate[\outer][\iter])^2 \left( \variance[\grad \obj] + \langmultmax\variance[\grad \constr] \right) + 2 \learnrate[\outer][\iter] \xi^{(\iter)}  \label{eq:left_off_convex}
\end{align}
}
Since $\val$ is convex in $\outer$ for all $\inner \in \innerset$ and $\langmult \in \langmults$, we then have $\left<\grad[\outer] \val(\outer[][][][\iter]) ,\left( \outer - \outer[][][][\iter] \right) \right> \leq  \val( \outer) - \val(\outer[][][][\iter])$, or re-organizing terms, $- \left<\grad[\outer] \val(\outer[][][][\iter]) ,\left( \outer[][][][\iter] - \outer \right) \right> \leq  - \left(\val(\outer[][][][\iter]) - \val(\outer) \right)$ giving us:

\begin{align}
    &\leq \Ex\left[ \left\| \outer[][][][\iter] - \outer \right\|^2 \right] - 2 \learnrate[\outer][\iter]  \left( \Ex\left[ \val(\outer[][][][\iter]) - \val(\outer) \right]\right) + (\learnrate[\outer][\iter])^2 \left( \variance[\grad \obj] + \langmultmax\variance[\grad \constr] \right) + 2 \learnrate[\outer][\iter]  \xi^{(\iter)}  \\
    &\leq \Ex\left[ \left\| \outer[][][][\iter] - \outer \right\|^2 \right] - 2 \learnrate[\outer][\iter] \left( \Ex\left[ \val(\outer[][][][\iter]) \right] - \val(\outer) \right) + (\learnrate[\outer][\iter])^2 \left( \variance[\grad \obj] + \langmultmax\variance[\grad \constr] \right) + 2 \learnrate[\outer][\iter] \xi^{(\iter)}   \enspace .
\end{align}

Taking $\outer \doteq \outer[][][*] \in \min_{\outer \in \outerset} \val(\outer)$, and unrolling the inequality from $\iter = \numiters[\outer]- 1$ to $\iter = 0$, each time conditioning on $\outer[][][][\iter]$ we get:
{\small
\begin{align}
    \Ex\left[ \left\| \outer[][][][\numiters] - \outer[][][*] \right\|^2 \right] &\leq \Ex\left[ \left\| \outer[][][][0] - \outer[][][*] \right\|^2 \right] - \sum_{\iter = 0}^{\numiters[\outer]-1} 2 \learnrate[\outer][\iter] \left( \Ex\left[ \val(\outer[][][][\iter]) \right] - \val(\outer[][][*]) \right) + \sum_{\iter = 0}^{\numiters[\outer]-1} (\learnrate[\outer][\iter])^2\left( \variance[\grad \obj] + \langmultmax\variance[\grad \constr] \right) + \sum_{\iter = 0}^{\numiters[\outer]-1} 2 \learnrate[\outer][\iter]   \xi^{(\iter)} 
\end{align}
\begin{align}
    \sum_{\iter = 0}^{\numiters[\outer]-1} 2 \learnrate[\outer][\iter]   \left( \Ex\left[ \val(\outer[][][][\iter]) \right] - \val(\outer[][][*]) \right) &\leq \Ex\left[ \left\| \outer[][][][0] - \outer[][][*] \right\|^2 \right]+ \sum_{\iter = 0}^{\numiters[\outer]-1} (\learnrate[\outer][\iter])^2 \left( \variance[\grad \obj] + \langmultmax\variance[\grad \constr] \right) + \sum_{\iter = 0}^{\numiters[\outer]-1} 2 \learnrate[\outer][\iter]  \xi^{(\iter)}   
\end{align}
}
Dividing both sides by $\left( \sum_{\iter = 0}^{\numiters[\outer]-1} \learnrate[\outer][\iter] \right)$, we obtain:
\begin{align}
    \sum_{\iter = 0}^{\numiters[\outer]-1}\learnrate[\outer][\iter]  \left( \Ex\left[ \val(\outer[][][][\iter]) \right] - \val(\outer[][][*]) \right) &\leq \Ex\left[ \left\| \outer[][][][0] - \outer[][][*] \right\|^2 \right]+ \sum_{\iter = 0}^{\numiters[\outer]-1} (\learnrate[\outer][\iter])^2 \left( \variance[\grad \obj] + \langmultmax\variance[\grad \constr] \right) + \sum_{\iter = 0}^{\numiters[\outer]-1} \learnrate[\outer][\iter] \xi^{(\iter)} \\
    \sum_{\iter = 0}^{\numiters[\outer]-1} \frac{\learnrate[\outer][\iter] \left( \Ex\left[ \val(\outer[][][][\iter]) \right]   - \val(\outer[][][*]) \right)}{\sum_{\iter = 0}^{\numiters[\outer]-1} \learnrate[\outer][\iter]} &\leq \frac{\Ex\left[ \left\| \outer[][][][0] - \outer[][][0] \right\|^2 \right] + \left( \variance[\grad \obj] + \langmultmax\variance[\grad \constr] \right) \sum_{\iter = 0}^{\numiters[\outer]-1} (\learnrate[\outer][\iter])^2 + \sum_{\iter = 0}^{\numiters[\outer]-1}\learnrate[\outer][\iter] \xi^{(\iter)} }{\sum_{\iter = 0}^{\numiters[\outer]-1} \learnrate[\outer][\iter] }\label{eq:left_off_conv_learnrate}
\end{align}

\paragraph{Fixed Learning Rate}
Let $\mean[\outer] \doteq \frac{\sum_{\iter = 0}^{\numiters[\outer] -1}  \learnrate[\outer][\iter]\outer[][][][\iter]}{\sum_{\iter = 0}^{\numiters[\outer] - 1}  \learnrate[\outer][\iter]}$, setting $\left\{\learnrate[\outer][\iter]\right\}_\iter \doteq \left\{\frac{1}{\max\{1, \lipschitz[\lang] \} }\right\}_\iter$ and using the convexity of $\val$, we then get:
\begin{align}
     \sum_{\iter = 0}^{\numiters[\outer]-1} \frac{\learnrate[\outer][\iter] \left( \Ex\left[ \val(\outer[][][][\iter]) \right]   - \val(\outer[][][*]) \right)}{\sum_{\iter = 0}^{\numiters[\outer]-1} \learnrate[\outer][\iter]} &\leq \frac{\Ex\left[ \left\| \outer[][][][0] - \outer[][][0] \right\|^2 \right] + \left( \variance[\grad \obj] + \langmultmax\variance[\grad \constr] \right) \sum_{\iter = 0}^{\numiters[\outer]-1} \learnrate[\outer][\iter] + \sum_{\iter = 0}^{\numiters[\outer]-1}\learnrate[\outer][\iter] \xi^{(\iter)} }{\sum_{\iter = 0}^{\numiters[\outer]-1} \learnrate[\outer][\iter] }\\
    \Ex\left[\val(\mean[\outer] )\right] - \val(\outer[][][*]) &\leq \frac{ \lipschitz[\lang] \left\| \outer[][][][0] - \outer[][][*][] \right\|^2 }{\numiters[\outer]}  +\left( \variance[\grad \obj] + \langmultmax\variance[\grad \constr] \right)  + \frac{\sum_{\iter = 0}^{\numiters[\outer] -1}  \learnrate[\outer][\iter] \xi^{(\iter)}}{\sum_{\iter = 0}^{\numiters[\outer] - 1}  \learnrate[\outer][\iter]}
\end{align}

\paragraph{Decreasing Learning rate}
Going back to \Cref{eq:left_off_conv_learnrate}, and instead setting a decreasing step size $\left\{\learnrate[\outer][\iter]\right\}_\iter \doteq \left\{\frac{1}{\sqrt{\iter + 1}}\right\}_\iter$ and using the convexity of $\val$, we instead get:
\begin{align}
    \Ex\left[\val(\mean[\outer] )\right] - \val(\outer[][][*]) &\leq \frac{ \left\| \outer[][][][0] - \outer[][][*][] \right\|^2 }{\numiters[\outer]}  + \frac{ \log(\numiters[\outer]) \left( \variance[\grad \obj] + \langmultmax\variance[\grad \constr] \right)}{ \sqrt{\numiters[\outer]} }  + \frac{\sum_{\iter = 0}^{\numiters[\outer] -1}  \learnrate[\outer][\iter] \xi^{(\iter)}}{\sum_{\iter = 0}^{\numiters[\outer] - 1}  \learnrate[\outer][\iter]}
\end{align}

Letting $\mean[{\xi}] \doteq \frac{\sum_{\iter = 0}^{\numiters[\outer] -1}  \learnrate[\outer][\iter] \xi^{(\iter)}}{\sum_{\iter = 0}^{\numiters[\outer] - 1}  \learnrate[\outer][\iter]}$, this implies that we converge to a $(\varepsilon + \mean[{\xi}], \delta)$-Stackelberg equilibrium in $\tilde{O}\left(\frac{\left\| \outer[][][][0] - \outer[][][0] \right\|^2 +  \left( \variance[\grad \obj] + \langmultmax\variance[\grad \constr] \right)}{\varepsilon^2}\right)$ gradient evaluations.



\paragraph{Strongly-Convex Case.}
Now suppose that the marginal function is strongly convex, which can be ensured by assuming in addition to \Cref{assum:convex_concave}, that for all parameters $\obji \in \objs$, $\obj[\obji]$ is strongly-convex in $\outer$. Picking up from \Cref{eq:left_off_convex}, and taking $\outer = \outer[][][][\iter]$ we have:

\begin{align}
     &\Ex\left[ \left\| \outer[][][][\iter + 1] - \outer \right\|^2 \right]\\
     &\Ex\left[  \left\| \outer[][][][\iter] - \outer \right\|^2 \right] - 2 \learnrate[\outer][\iter]   \Ex\left[ \left<\grad \val(\outer[][][][\iter]), \outer[][][][\iter] - \outer \right>\right]  + (\learnrate[\outer][\iter])^2 \left( \variance[\grad \obj] + \langmultmax\variance[\grad \constr] \right) + 2 \learnrate[\outer][\iter]   \xi^{(\iter)}\\
     &\leq \Ex\left[  \left\| \outer[][][][\iter] - \outer \right\|^2 \right] - 2 \learnrate[\outer][\iter]  \Ex\left[ \frac{\scparam[\outer]}{2} \left\|\outer[][][][\iter] - \outer \right\|^2 + \val(\outer[][][][\iter]) - \val(\outer) \right]  + (\learnrate[\outer][\iter])^2 \left( \variance[\grad \obj] + \langmultmax\variance[\grad \constr] \right) + 2 \learnrate[\outer][\iter]   \xi^{(\iter)}\\
     &\leq \Ex\left[  \left\| \outer[][][][\iter] - \outer \right\|^2 \right] -  \learnrate[\outer][\iter] \scparam[\outer] \Ex\left[ \left\|\outer[][][][\iter] - \outer \right\|^2\right] + 2 \learnrate[\outer][\iter] \Ex\left[  \val(\outer[][][][\iter]) - \lang(\outer) \right]  + (\learnrate[\outer][\iter])^2 \left( \variance[\grad \obj] + \langmultmax\variance[\grad \constr] \right) + 2 \learnrate[\outer][\iter]   \xi^{(\iter)}\\
     &\leq \left( 1-  \learnrate[\outer][\iter] \scparam[\outer] \right) \Ex\left[ \left\|\outer[][][][\iter] - \outer \right\|^2\right] - 2 \learnrate[\outer][\iter] \Ex\left[  \val(\outer[][][][\iter]) - \val(\outer) \right]  + (\learnrate[\outer][\iter])^2 \left( \variance[\grad \obj] + \langmultmax\variance[\grad \constr] \right) + 2 \learnrate[\outer][\iter]   \xi^{(\iter)}
\end{align}
where the second line follows from the $\scparam[\outer]$-strong convexity of $\val$. 

Taking $\outer \doteq \outer[][][*] \in \argmin_{\outer \in \outerset} \val(\outer)$, and re-organizing expressions:
{\small
\begin{align}
    2\learnrate[\outer][\iter]\left( \Ex\left[ \val(\outer[][][][\iter]) \right]  - \val(\outer[][][*]) \right) &\leq \left( 1-  \learnrate[\outer][\iter] \scparam[\outer] \right) \Ex\left[ \left\|\outer[][][][\iter] - \outer[][][*] \right\|^2\right] - \Ex\left[ \left\| \outer[][][][\iter + 1] - \outer[][][*] \right\|^2 \right] + (\learnrate[\outer][\iter])^2 \left( \variance[\grad \obj] + \langmultmax\variance[\grad \constr] \right) + 2 \learnrate[\outer][\iter]  \xi^{(\iter)}
\end{align}
}
Restricting for all $\iter \in \N_+$, $\learnrate[\outer][\iter] \leq 1$ and dividing the right hand side by $\learnrate[\outer][\iter]$ we then have:
{\small
\begin{align}
    2 \learnrate[\outer][\iter]\left( \Ex\left[ \val(\outer[][][][\iter]) \right]  - \val(\outer) \right) &\leq \frac{\left( 1-  \learnrate[\outer][\iter] \scparam[\outer] \right)}{\learnrate[\outer][\iter]}\Ex\left[ \left\|\outer[][][][\iter] - \outer \right\|^2\right] - \frac{1}{\learnrate[\outer][\iter]} \Ex\left[ \left\| \outer[][][][\iter + 1] - \outer \right\|^2 \right] + \learnrate[\outer][\iter] \left( \variance[\grad \obj] + \langmultmax\variance[\grad \constr] \right) + 2\xi^{(\iter)} \\
    \learnrate[\outer][\iter]\left( \Ex\left[ \val(\outer[][][][\iter]) \right]  - \val(\outer) \right) &\leq \frac{\left( 1-  \learnrate[\outer][\iter] \scparam[\outer] \right)}{2 \learnrate[\outer][\iter]}\Ex\left[ \left\|\outer[][][][\iter] - \outer \right\|^2\right] - \frac{1}{2 \learnrate[\outer][\iter]} \Ex\left[ \left\| \outer[][][][\iter + 1] - \outer \right\|^2 \right] + \left( \variance[\grad \obj] + \langmultmax\variance[\grad \constr] \right) + \learnrate[\outer][\iter] \xi^{(\iter)}   
\end{align}
}

Setting for all $\iter \in \N_+$, $\learnrate[\outer][\iter] \doteq \frac{2}{\scparam(\iter + 1)}$, and substituting accordingly on the right hand side, we then obtain:
\begin{align}
    \learnrate[\outer][\iter] \left(\Ex\left[ \val(\outer[][][][\iter]) \right]  - \val(\outer) \right) &\leq  \frac{\scparam[\outer]\left( \iter - 1  \right)}{4} \Ex\left[ \left\|\outer[][][][\iter] - \outer \right\|^2\right] - \frac{\scparam[\outer]\left( \iter + 1  \right)}{4} \Ex\left[ \left\|\outer[][][][\iter + 1] - \outer \right\|^2\right] + \left( \variance[\grad \obj] +  \langmultmax\variance[\grad \constr] \right) +  \learnrate[\outer][\iter] \xi^{(\iter)}
\end{align}

Unrolling the inequality from $\iter = 0$ to $\iter = \numiters[\outer]- 1$, each time conditioning on $\outer[][][][\iter]$ we get:
{\small
\begin{align}
    &\sum_{\iter = 0}^{\numiters[\outer]-1} \learnrate[\outer][\iter] \left(\Ex\left[ \val(\outer[][][][\iter]) \right]  - \val(\outer[][][*]) \right)\\
    &\leq  \sum_{\iter = 0}^{\numiters[\outer]-1} \frac{\scparam[\outer]\left( \iter - 1  \right)}{4} \Ex\left[ \left\|\outer[][][][\iter] - \outer[][][*] \right\|^2\right] - \sum_{\iter = 0}^{\numiters[\outer]-1} \frac{\scparam[\outer]\left( \iter + 1  \right)}{4} \Ex\left[ \left\|\outer[][][][\iter + 1] - \outer[][][*] \right\|^2\right] + \numiters[\outer]\left( \variance[\grad \obj] + \langmultmax\variance[\grad \constr] \right) + \sum_{\iter = 0}^{\numiters[\outer]-1} \learnrate[\outer][\iter] \xi^{(\iter)}\\
    &\leq  - \frac{\scparam[\outer]\numiters\left( \numiters[\outer]+ 1 \right)}{4} \Ex\left[ \left\|\outer[][][][\iter + 1] - \outer[][][*] \right\|^2\right] + \numiters[\outer]\left( \variance[\grad \obj] + \langmultmax\variance[\grad \constr] \right) + \sum_{\iter = 0}^{\numiters[\outer]-1} \learnrate[\outer][\iter] \xi^{(\iter)}\\
    &\leq \numiters[\outer]\left( \variance[\grad \obj] + \langmultmax\variance[\grad \constr] \right) + \sum_{\iter = 0}^{\numiters[\outer]-1}  \learnrate[\outer][\iter] \xi^{(\iter)}
\end{align}
}
Let $\mean[\outer] \doteq \frac{\sum_{\iter = 0}^{\numiters[\outer] -1}  \learnrate[\outer][\iter] \outer[][][][\iter]}{\sum_{\iter = 0}^{\numiters[\outer] - 1} \learnrate[\outer][\iter]}$. Multiplying both sides by $\frac{1}{\sum_{\iter = 0}^{\numiters[\outer]-1} \learnrate[\outer][\iter]} = \frac{1}{\sum_{\iter = 0}^{\numiters[\outer]-1} \frac{\scparam}{2(\iter + 1)}} \leq \frac{\scparam}{2\numiters[\outer] (\numiters[\outer] + 1) } $:

\begin{align}
     \frac{\sum_{\iter = 0}^{\numiters[\outer]-1} \learnrate[\outer][\iter] \left(\Ex\left[ \val(\outer[][][][\iter]) \right]  - \val(\outer[][][*]) \right)}{\sum_{\iter = 0}^{\numiters[\outer]-1} \learnrate[\outer][\iter]} &\leq \frac{\scparam \numiters[\outer]\left( \variance[\grad \obj] + \langmultmax\variance[\grad \constr] \right)}{2\numiters[\outer] (\numiters[\outer] + 1)} + \frac{\sum_{\iter = 0}^{\numiters[\outer]-1}  \learnrate[\outer][\iter] \xi^{(\iter)}}{\sum_{\iter = 0}^{\numiters[\outer]-1} \learnrate[\outer][\iter]}\\
     \Ex\left[ \val(\mean[\outer]) \right]  - \val(\outer)  &\leq \frac{\scparam \left( \variance[\grad \obj] + \langmultmax\variance[\grad \constr] \right)}{\numiters[\outer]+ 1} + \frac{\sum_{\iter = 0}^{\numiters[\outer]-1} \learnrate[\outer][\iter] \xi^{(\iter)}}{\sum_{\iter = 0}^{\numiters[\outer]-1} \learnrate[\outer][\iter]}
\end{align}

Letting $\mean[{\xi}] \doteq \frac{\sum_{\iter = 0}^{\numiters[\outer] -1}  \learnrate[\outer][\iter] \xi^{(\iter)}}{\sum_{\iter = 0}^{\numiters[\outer] - 1}  \learnrate[\outer][\iter]}$, this implies that we converge to a $(\varepsilon + \mean[{\xi}], \delta)$-Stackelberg equilibrium in in $O\left(\frac{\scparam\left( \variance[\grad \obj] + \langmultmax\variance[\grad \constr] \right)}{\varepsilon}\right)$ gradient evaluations. 


\end{proof}

\subsection{Sufficient Conditions Markov Stackelberg Games}

\begin{restatable}[Concavity of the state-value function for bilinear parameterizations]{lemma}{lemmaconvexityparamsfollower}
\label{lemma:concavity_params_follower}
Fix a leader policy $\outerpoint \in \outerset$.
For any zero-sum Markov Stackelberg game $\game$, the state-value function $\statevalue[{\outerpoint, \inner}] (\state)$ is concave in $\inner$, for all $\state \in \states$ if:
    1.~the state and action spaces $\states$ and $\inneractions$ are non-empty, compact, and convex;
    2.~for all leader actions $\outeraction \in \outeractions$, the reward function $(\state, \inneraction) \mapsto \reward(\state, \outeraction, \inneraction)$ is continuous, concave, and bounded;
    3.~for all leader actions $\outeraction \in \outeractions$, the transition function $(\state, \inneraction) \mapsto \trans(\cdot \mid \state, \outeraction, \inneraction)$ is continuous and stochastically concave;
    4.~the follower's policy $(\inner, \state) \mapsto \policy[\inner] (\state)$ is bilinear, and the leader's policy $\state \mapsto \policy[\outerpoint](\state)$ is affine.
%
\end{restatable}

\begin{remark}
For zero-sum Markov (simultaneous-move) games (resp. MDPs) with continuous state and action spaces, this lemma provides sufficient conditions for the state-value function to be convex-concave (resp. concave) in the policy parameters. 
Additionally, this result implies that for infinite horizon discounted linear quadratic regulators (LQR) \cite{fazel2018global}, and linear MDPs \cite{bradtke1996linear, melo2007q, jin2020provably}) with a concave feature map, the state-value function is concave in the policy parameters. Note that for LQRs the bilinear policy parametrization is optimal \cite{bertsekas2012dynamic}.

\end{remark}

Next, we provide conditions which guarantee that the marginal function $\marginalfunc$ is convex in $\outer$.

\begin{restatable}[Convexity of the marginal function]{lemma}{lemmaconvexityparamsleader}
\label{lemma:convexity_params_leader}
    For any zero-sum Markov Stackelberg game $\game$, the marginal function $\marginalfunc[][](\outer)$ is convex in $\outer$, for all $\state \in \states$, if:
    1.~the state and action spaces $\states$ and $\inneractions$ are non-empty, compact, and convex;
    2.~the action correspondence $\outeraction \mapsto \coupledactions(\state, \outeraction)$ is concave in $\outeraction$ and convex-valued, for all $\state \in \states$; 
    3.~the reward function $\reward$ is continuous, convex,\amy{in what?}\deni{in everything!} and bounded, i.e., $\left\|\reward \right\|_\infty < \infty$;
    4.~$\trans(\cdot \mid \state, \outeraction, \inneraction)$ is continuous and stochastically convex \sdeni{}{in $\state, \outeraction, \inneraction$} \amy{in what?}\deni{in everything!},
    5.~the policies $(\outer, \state) \mapsto \policy[\outer](\state)$ and $ (\inner, \state) \mapsto \policy[\inner](\state)$ are bilinear.
\end{restatable}

\lemmaconvexityparamsfollower*

\begin{proof}[Proof of \Cref{lemma:convexity_params_leader}]
    Recall that the action-value function $\actionvalue[{\outerpoint, \inner}]$ can be written as \cite{bellman1952theory}:
    \begin{align}
        \actionvalue[{\outerpoint, \inner}](\state, \outeraction, \inneraction) = \reward(\state, \outeraction, \inneraction) + \discount\Ex_{\staterv \sim \trans( \cdot \mid \state, \outeraction, \inneraction)}\left[\statevalue[{\outerpoint, \inner}](\staterv)\right]
    \end{align}

    Suppose that $\statevalue[{\outerpoint, \inner}](\state)$ is continuous, bounded, and concave in $\state$.
    Let $\inneraction, \inneraction[][][\prime] \in \inneractions$, and $\lambda \in (0,1)$. Since $\trans(\cdot \mid \state, \outeraction, \inneraction)$ is CSD concave in $\state$ and in $\inneraction$, we have:
    \begin{align}
        \Ex_{\staterv \sim \trans( \cdot \mid \state, \outeraction, \lambda \inneraction + (1-\lambda)\inneraction[][][\prime])}\left[\statevalue[{\outerpoint, \inner}](\staterv)\right] &\geq \Ex_{\staterv \sim \lambda\trans( \cdot \mid \state, \outeraction,  \inneraction) + (1-\lambda) \trans( \cdot \mid \state, \outeraction,  \inneraction[][][\prime])}\left[\statevalue[{\outerpoint, \inner}](\staterv)\right]\\
        &\geq \lambda  \Ex_{\staterv \sim \trans( \cdot \mid \state, \outeraction, \inneraction )}\left[\statevalue[{\outerpoint, \inner}](\staterv)\right] + (1-\lambda) \Ex_{\staterv \sim \trans( \cdot \mid \state, \outeraction,  \inneraction[][][\prime])}\left[\statevalue[{\outerpoint, \inner}](\staterv)\right]
    \end{align}

    Hence, $\Ex_{\staterv \sim \trans( \cdot \mid \state, \outeraction, \inneraction)}\left[\statevalue[{\outerpoint, \inner}](\staterv)\right]$ is concave in $\inneraction$. As a result, since for all states $\state \in \states$, $\reward(\state, \outeraction, \inneraction)$ is concave in $\inneraction$, $\actionvalue[{\outerpoint, \inner}](\state, \outeraction, \inneraction)$ is also concave in $\state$ and in $\inneraction$. 
    A similar argument holds for the concavity of $\actionvalue[{\outerpoint, \inner}](\state, \outeraction, \inneraction)$ in $\state$.
    
    Additionally, as $\reward$ and $\statevalue[{\outerpoint, \inner}]$ are continuous and bounded, $\actionvalue[{\outerpoint, \inner}](\state, \outeraction, \inneraction)$ is also continuous and bounded. 
    
    Furthermore, since $\actionvalue[{\outerpoint, \inner}](\state, \outeraction, \inneraction)$ is continuous, bounded, concave in $\state$, and in $\inneraction$, its composition with $(\state, \inner) \mapsto (\state, \outeraction, \policy[\inner](\state))$, i.e., $\actionvalue[{\outerpoint, \inner}](\state, \outeraction, \policy[\inner](\state))$, is also continuous, bounded, as well as concave in $\state$ and in $\inner$ since it is the composition of a concave function in $\state$ and in $\inneraction$ with an affine function in $\state$ and in $\inner$. 

    Next, we will prove for  that $\statevalue[{\outerpoint, \inner}]$ is  continuous, bounded, and concave in $\state$.
    
    Let $\calC$ be the space of all continuous and bounded functions on $[\nicefrac{\rewardbound}{1-\discount}, \nicefrac{\rewardbound}{1-\discount}]$. 
    Define the Bellman policy operator $\bellpol[{(\outerpoint, \inner)}]: \calC \to \calC$:
    \begin{align}
        \bellpol[{(\outerpoint, \inner)}] (\statevalue) \doteq \reward(\state, \policy[\outerpoint](\state), \policy[\inner](\state)) + \discount\Ex_{\staterv \sim \trans( \cdot \mid \state, \policy[\outerpoint](\state), \policy[\inner](\state))}\left[\statevalue(\staterv)\right]
    \end{align}
    Recall that $\statevalue[{\outerpoint, \inner}]$ is the unique fixed point of the Bellman policy operator, i.e., $\bellpol[{(\outerpoint, \inner)}](\statevalue[{\outerpoint, \inner}]) = \statevalue[{\outerpoint, \inner}]$ \cite{bellman1952theory}.
    
    Let $\calC^{\mathrm{concave}}$ be the space of all continuous, bounded, and concave functions on $[\nicefrac{\rewardbound}{1-\discount}, \nicefrac{\rewardbound}{1-\discount}]$. 
    Since $\actionvalue[{\outerpoint, \inner}](\state, \outeraction, \policy[\inner](\state)) = \reward(\state, \policy[\outer](\state), \policy[\inner](\state)) + \discount\Ex_{\staterv \sim \trans( \cdot \mid \state, \policy[\outer](\state), \policy[\inner](\state))}\left[\statevalue[{\outerpoint, \inner}](\staterv)\right]$ is concave in $\state$, the Bellman policy operator $\bellpol[{(\outerpoint, \inner)}]$ preserves concavity. Hence, we can restrict the Bellman policy operator to $\calC^{\mathrm{concave}}$ s.t. $\bellpol[{(\outerpoint, \inner)}]|_{\calC^{\mathrm{concave}}}: \calC^{\mathrm{concave}} \to \calC^{\mathrm{concave}}$.
    Then, by the Banach fixed point theorem \cite{banach1922operations}, $\bellpol[{(\outerpoint, \inner)}] (\statevalue) \doteq \reward(\state, \policy[\outer](\state), \policy[\inner](\state)) + \discount\Ex_{\staterv \sim \trans( \cdot \mid \state, \policy[\outer](\state), \policy[\inner](\state))}\left[\statevalue[{\outerpoint, \inner}](\staterv)\right]$, since the restricted Bellman policy operator is a contraction mapping, the restricted Bellman policy operator $\bellpol[{(\outerpoint, \inner)}]|_{\calC^{\mathrm{concave}}}$ has a unique fixed point, i.e., $\bellpol[{(\outerpoint, \inner)}](\statevalue[{\outerpoint, \inner}]) = \statevalue[{\outerpoint, \inner}]$. As $\calC^{\mathrm{concave}} \subset \calC$, its fixed points must coincide with the fixed points of $\bellpol[{(\outerpoint, \inner)}]$. As such $\statevalue[{\outerpoint, \inner}](\state)$ must be concave in $\state$ which in turn implies $\statevalue[{\outerpoint, \inner}](\state)$ must be continuous, bounded and concave in $\state$ and in $\inner$.

\end{proof}


\begin{proof}[Proof of \Cref{lemma:concavity_params_follower}]
    We can apply \Cref{lemma:convexity_params_leader} to $(\outer, \inner)$ to deduce that if 
    \begin{enumerate}
        \item the state space $\states$ is continuous, non-empty, compact, and convex; 
        \item the payoffs $\reward$ are bounded, i.e., $\left\|\reward \right\|_\infty < \infty$;
        \item for all states $\state \in \states$, $\reward(\state, \outeraction, \inneraction)$ is continuous and convex in both $\state$ and $(\outeraction, \inneraction)$;
        \item $\trans(\cdot \mid \state, \outeraction, \inneraction)$ is continuous and stochastically convex in $\state$, and continuous and stochastically convex in $(\outeraction, \inneraction)$,
        \item the policy $\policy[\inner](\state)$ is affine in $\inner$ and in $\state$, i.e., it is bilinear.
    \end{enumerate}
    $\statevalue[\outer, \inner]$ is convex in $(\outer, \inner)$.

    Then, if in addition, we assume that the action correspondence $\coupledactions$ is concave, \deni{product of concave correspondences is concave and composition with affine function is concave.} by \citet{fiacco1986convexity}, we have that for any state $\state \in \states$, $\marginalfunc[][{\policy[\outeraction]}] (\state) = \max_{\policy[\inneraction] \in \coupledactions(\policy[\outeraction])} \statevalue[{\policy[\outeraction], \policy[\inneraction]}](\state)$ is convex.
\end{proof}

\thmconvpgda*

\begin{proof}[Proof of \Cref{thm:conv_pgda}]
    First of all note that, the REINFORCE estimator is an unbiased estimator of the gradient. Additionally, under the assumptions of \Cref{lemma:concavity_params_follower} and \Cref{lemma:convexity_params_leader}, the policies satisfy the following:
         $\left\| \grad[\outer] \log(\policy[\outer](\state))\right\| < \infty
         \left\| \grad[\inner]\log(\policy[\inner](\state)) \right\| < \infty
         \left\| \grad[\outer][2] \log(\policy[\outer](\state))\right\| < \infty
         \left\| \grad[\inner][2] \log(\policy[\inner](\state)) \right\| < \infty$.
    Hence, by Lemma B.2 of \citeauthor{papini2018stochastic} \cite{papini2018stochastic}, we must have that $\cumulutil$ is Lipschitz-smooth. Additionally, by Lemma 4.1 of \citeauthor{shen2019hessian} \cite{shen2019hessian}, the REINFORCE estimator has an expected squared Euclidean norm which is bounded, i.e., bounded variance.

    As such, under \Cref{assum:smooth_stoch} and the assumptions of \Cref{lemma:concavity_params_follower} and \Cref{lemma:convexity_params_leader}, the conditions for \Cref{thm:min_max_convergence} hold, and nested policy GDA converges to a Stackelberg equilibrium of
    $\min_{\outer \in \outerset} \max_{\inner \in \coupledactions (\outer)} \cumulutil(\outer, \inner)$.

    This convergence rate then implies convergence to a recursive Stackelberg equilibrium simply by noting that for any distribution over the state space $\rho \in \simplex(\states)$, the following relationship defines the distribution mismatch coefficient (see, for instance, Lemma 4.1 of \citeauthor{agarwal2020optimality}\cite{agarwal2020optimality}):
    \begin{align}
        \min_{\state \in \states} \left\{ \frac{\statedist[{\initstates}][(\outer, \inner)](\state)}{\statedist[{\rho}][(\outer, \inner)](\state)} \right\} \statevalue[\outer, \inner](\rho) \leq \cumulutil(\outer, \inner) \leq \max_{\state \in \states} \left\{ \frac{\statedist[{\initstates}][(\outer, \inner)](\state)}{\statedist[{\rho}][(\outer, \inner)](\state)} \right\} \statevalue[\outer, \inner](\rho)
    \end{align}
\end{proof}

\thmreachavoid*

\begin{proof}[Proof of \Cref{thm:reach_avoid}]
    Since $\detertrans$ is affine in $\state$ and $(\outeraction, \inneraction)$, it is continuous and both stochastically convex and stochastically concave (see Proposition 1 of \citeauthor{atakan2003valfunc} \cite{atakan2003valfunc}). 
    Additionally, notice that \amy{which $\reward$? there are two? need two arguments here!} $\reward$ is continuous, bounded, convex in $\state[][\outeraction]$, and concave in $\state[][\inneraction]$. 
    Since policies are stochastic, expected rewards depend only on $\inneraction$, \amy{does not follow from policies being stochastic?} and hence are jointly convex in $(\outeraction, \inneraction)$ and concave in $\inneraction$.
    Furthermore, as $\coupledactions$ is concave, and policies are bilinear, the assumptions of \Cref{lemma:concavity_params_follower} and \Cref{lemma:convexity_params_leader} are satisfied.
    \sdeni{}{(We note that since the state of the defender and attacker are separate, the assumptions required by the lemmas on the state spaces apply separately for both players.)}
    Finally, notice that \amy{i didn't notice; please explain.} the assumptions of \Cref{assum:smooth_stoch} hold, and hence \Cref{thm:conv_pgda} applies.
    \label{proof:reach-avoid}
\end{proof}

Sufficient conditions: En route to establishing conditions for convex-concavity, let us fix the leader's policy $\outerpoint \in \outerset$, in which case the follower's optimization problem becomes an MDP.
A large body of work \cite{agarwal2020optimality, agarwal2021theory, kakade2002approximately, zhang2020global} has been dedicated to understanding properties of $\cumulutil(\outerpoint, \inner)$. 
For MDPs with finite state and action spaces, it is known that $\cumulutil(\outerpoint, \inner)$ is incave in $\inner$ (or satisfies a gradient dominance condition), i.e., its \sdeni{local maxima}{stationary points} correspond to global maxima.
Beyond finite state and action space MDPs, however, it is not known how to guarantee the incavity of $\cumulutil(\outerpoint, \inner)$ in $\inner$.
We present conditions in \Cref{lemma:concavity_params_follower} that ensure concavity, not incavity, and hence a stronger property. 


\newpage
\section{Experiments Setup}
\paragraph{Bellman Error}
The Bellman error is defined as:
\begin{equation}
    \mathop{\Ex}_{\staterv \sim \mu} \left[ \left| \mathop{\Ex}_{\histrv \sim \histdistrib[][\policy]} \left[ \actionvalue[][{(\outer, \inner)}] (\staterv, \policy[\outer] (\staterv), \policy[\inner] (\staterv)) \right] - \mathop{\Ex}_{\histrv \sim \histdistrib[][\policy]} \left[ \statevalue[][{(\outer, \inner)}] (\staterv) \right] \right| \right]
\label{eq:bellman}
\end{equation}

We use Monte Carlo simulations to estimate the expectations in this error empirically.
Specifically, we sum over rolled out trajectories, given the policy profile $(\outer, \inner)$.
Given a history $\hist[][][] = (\state[\iter], \outeraction[][][][\iter], \inneraction[][][][\iter])_{\iter = 0}w^{\numiters-1}$, it is straightforward to look up \amy{in our neural network!?} \sarjun{}{No, they are estimated by rolling out trajectories and calculating the payoff matrix through samples}the value $\statevalue[][{(\outer, \inner)}] (\state)$, for some state $\state \in \states$ by sdd.
The action values (i.e., and worthwhile deviations), however, must be calculated.

In the Nash case, \amy{is this correct? does it make sense to allow your local actions to change to something better, but then to evaluate using the same old next state $\state[][][']$? it seems to me that that should be an expectation over next states instead, since the local actions have changed?} \sarjun{I think this should have been an LP}{}
\begin{equation}
\label{eq:nash_error}
\actionvalue[{\text{\scriptsize Nash}}][{(\outer, \inner)}] (\state[][][(t)], \policy[\outer] (\state), \policy[\inner] (\state)) = \min_{\outeraction[][][\prime] \in \outeractions} \max_{\inneraction[][][\prime] \in \inneractions} \reward (\state, \outeraction[][][\prime], \inneraction[][][\prime]) + \gamma \statevalue[][{(\outer, \inner)}] (\state[][][(t+1)])
\end{equation}
\amy{i also worry that this is a seq'l expression. i mean, it looks like the follower has an advantage in this setup, because he moves second. is this convex-concave or something so that the order doesn't matter? i doubt it with a neural net, so changing the order of the decision making might change the value of \Cref{eq:nash_error}.}

\amy{macro question for Deni: why is Nash not a subscript in $\actionvalue[{\text{\scriptsize Nash}}][{(\outer, \inner)}]$?}

\samy{}{
In the Stackelberg case,
\begin{equation}
\label{eq:stackelberg_error}
\actionvalue[{\text{\scriptsize Stackelberg}}][{(\outer, \inner)}] (\state[][][(t)], \policy[\outer] (\state), \policy[\inner] (\state)) = \min_{\outeraction[][][\prime] \in \outeractions} \max_{\inneraction[][][\prime] \in \coupledactions (\state, \outeraction[][][\prime])} \reward (\state, \outeraction[][][\prime], \inneraction[][][\prime]) + \gamma \statevalue[][{(\outer, \inner)}] (\state[][][(t+1)])
\end{equation}
}
\amy{is this kosher? do these alternative actions $\outeraction[][][\prime]$ and $\inneraction[][][\prime]$ have to be representable by our neural network? (this question also applies in the Nash case.) if so, then $\inneractions, \outeractions$, and $\coupledactions (\state, \outeraction[][][\prime])$ all need to be described using parameterized language.}

In the Stackelberg variant $q^{\policy[\outer], \policy[\inner]} (\staterv, \outeraction, \inneraction)$ is verified \amy{verified? what does this mean?} by \Cref{algo:bellman}. \sarjun{}{the payoffs are estimated by sampling trajectories, so this is a way of empirically checking the bellman error}




The inner expectations of \Cref{eq:bellman} correspond to trajectories $\histrv$ rolled out under the policies $\policy[\outer]$ and $\policy[\inner]$ from the same starting state. The absolute value is taken, to make sure that high positive and negative errors do not average out. Finally, we take the outer expectation over starting states $\staterv$ drawn from the starting distribution $\mu$. 

\sarjun{}{We sample 32 states with 32 trajectories \amy{do you mean 32 action profiles instead of 32 trajectories?}, and apply this procedure five times, each with a different random seed.}

\begin{algorithm}
\caption{Stackelberg Equilibrium Verification \cite{conitzerStackelbergMixedStrategies2016}}
\label{alg:defender_strategy}
\begin{algorithmic}[1]
\State $\operatorname{maximize} \sum_{\outeraction \in \outeractions} q^{\policy[\outer], \policy[\inner]}(\state,\outeraction,\inneraction[][][][*]) p(\outeraction)$
\State subject to
\State $(\forall \inneraction \in \inneractions) \sum_{\outeraction \in \outeractions} \bigg( q^{\policy[\outer], \policy[\inner]}(\state,\outeraction,\inneraction[][][][*]) - q^{\policy[\outer], \policy[\inner]}(\state,\outeraction,\inneraction) \bigg) p(\outeraction) \geq 0 $
\State $\sum_{\outeraction \in \outeractions} p(\outeraction) = 1$ 
\State $(\forall \outeraction \in \outeraction) p(\outeraction) \geq 0$
\end{algorithmic}
\label{algo:bellman}
\end{algorithm}

\begin{figure}[htpb]
  \centering
    \includegraphics[width=\textwidth/2]{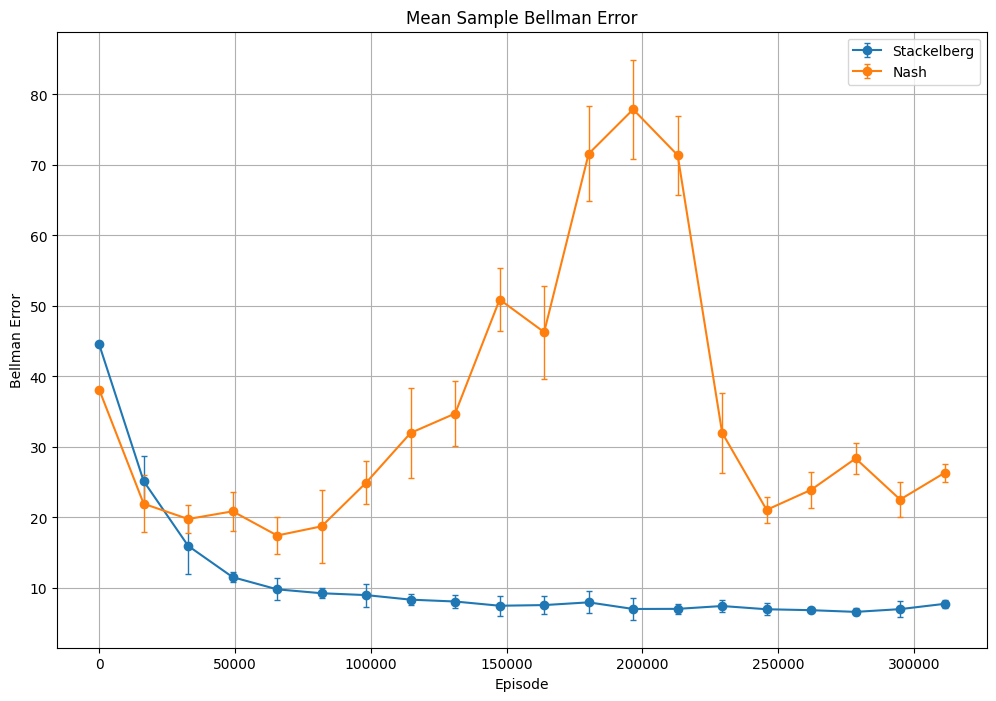}
  \caption{The average Bellman error of the Stackelberg version converges close to zero compared the the average Bellman error of the Nash}
  \label{fig:bellman}
\end{figure}

\Cref{fig:bellman} shows the Bellman errors we obtained by running \Cref{alg:nested_pgda} on the two aforementioned Stackelberg games, and \Cref{alg:sim_pgda} on the one simultaneous-move game.
The Bellman errors for the Stackelberg varient approach zero while Nash remains high indicating that the Stackelberg variant converged to an equilibrium. \amy{ok. but WHY doesn't Nash converge? which assumptions required of sim GDA are not satisfied here?}

\paragraph{Evaluation}
For our evaluation we play our attacker against a hand-crafted defender policy which takes the action which will minimize the distance to the attacker: $\outeraction = \argmin_{\outeraction} || \state[][\outeraction]['] - \state[][\inneraction]||_2$

While simple, this policy is still challenging and the optimal strategy in classical pursuit-evasion \cite{OylerPE}. This is made even more difficult by the fact that the defender spawns in between the attacker and $\targetstates$. In order to win, the attacker needs to execute a swerve maneuver \cite{isaacs1954differential} to beat the defender on its way to the goal. 

We test our learned attacker policy against this defender with 100 different random seeds, sampling $\staterv ~ \mu$. We find that the Nash policy is never able to beat this defender while the Stackelberg variant wins more than half the time.  

\begin{center}
\begin{tabular}{|l|l|l|l|}
\hline
 & Reached & Collision \\
\hline
Stackelberg & 62 & 38 \\
\hline
Nash & 0 & 100 \\
\hline
\end{tabular}
\label{tab:eval_game}
\end{center}

\begin{figure}[htpb]
    \centering
    \begin{subfigure}[t]{0.50\textwidth}
    \includegraphics[width=\textwidth]{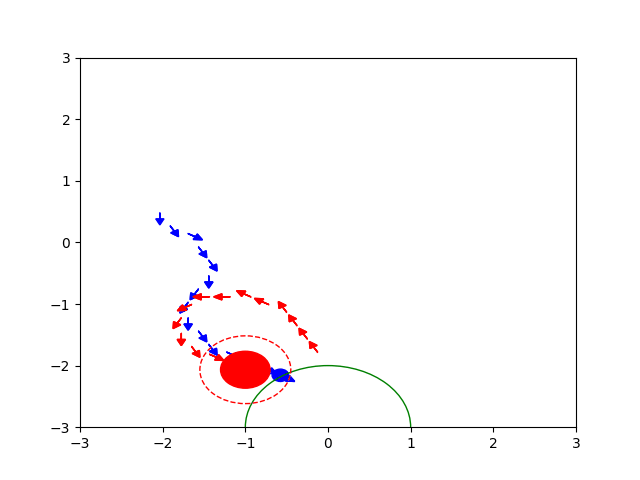}
        \caption{Stackelberg}
    \end{subfigure}
    \hfill 
    \begin{subfigure}[t]{0.49\textwidth}
\includegraphics[width=\textwidth]{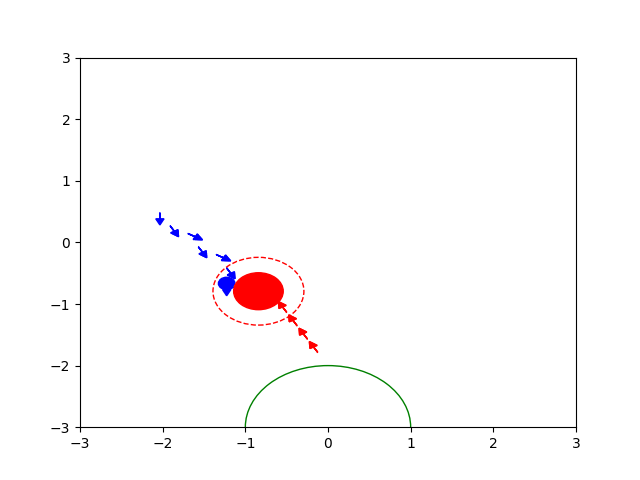}
        \caption{Nash}
    \end{subfigure}
    \caption{Both figures show the an example evaluation run. The Stackelberg policy is able to successfully execute the swerve  while the Nash is not.}
    \label{fig:eval_game}
\end{figure}

These results show that the Stackelberg attacker outperforms the Nash attacker by a large margin. This is likely due to two reasons. First, \Cref{alg:nested_pgda} appears to reach an equilibrium indicating that both the attacker and defender have learned reasonable policies. Second, masking appears to be helpful for learning safer policies \cite{huangCloserLookInvalid2022}. In particular, masking helps the attacker explore since an episode that would have otherwise terminated (had the attacker taken the illegal action), is able to continue.

\subsection{Training Setup}

The environment is an 7-by-7 continuous space with the goal at the bottom, ranging from $-1$ to $1$ along the $x$-axis and $-3$ to $-3$ along the y-axis. 
We make a small modification to the displacement function to handle the edge case where a player hits the wall.
If this happens, they bounce back in like a ray of light hitting a mirror: 
\begin{align}
\detertrans[{\w}] (\state[][{\w}], \w ) = 
v \w
\begin{bmatrix}
\cos( - \omega) & -\sin(- \omega) \\
\sin(- \omega) & \cos(- \omega) \\
\cos(\pi) & -\sin( - \omega) \\
\sin(- \omega) & \cos( - \omega) 
\end{bmatrix}
\begin{bmatrix}
x_{\text{boundary}} \\
y_1 \\
x_{\text{boundary}} \\
y_1
\end{bmatrix} \text{,  if $x$-boundary is violated }
\label{eq:bouncex} \\
\detertrans[{\w}] (\state[][{\w}], \w) = 
v \w
\begin{bmatrix}
\cos(\pi - \omega) & -\sin(\pi - \omega) \\
\sin(\pi - \omega) & \cos(\pi - \omega) \\
\cos(\pi) & -\sin(\pi - \omega) \\
\sin(\pi - \omega) & \cos(\pi - \omega) 
\end{bmatrix}
\begin{bmatrix}
x_1 \\
y_{\text{boundary}} \\
x_2 \\
y_{\text{boundary}}
\end{bmatrix} \text{,  if $y$-boundary is violated } 
\label{eq:bouncey}
\end{align}
Should this lead to a situation where the attacker has no legal moves available, the episode terminates. 


\begin{table}[htbp]
\centering
\caption{Training Parameters}
\begin{tabular}{|c|c|}
\hline
\textbf{Parameter} & \textbf{Value} \\ \hline\hline
Environment Min & $-3$ \\ \hline
Environment Max & 3 \\ \hline
Goal Position $g$ & (0, $-3$) \\ \hline
Defender Initial State Distribution $\initstates$& 
 $[0,-2,\text{up}]$\\ \hline
Attacker Starting States Distribution$\initstates$ & $U([-3, 3] \times [0,3] \times \text{down})$  \\ \hline
Goal Radius ($\ball[g][.]$) & 1 \\ \hline
Capture Radius ($\avoidstates$) & 0.3 \\ \hline
Velocity ($v$) & 0.25 \\ \hline
Turning angle ($\omega$) & $30^\circ$ \\ \hline
Max steps & 50 \\ \hline
Algorithm & REINFORCE with baseline (state-value function) \\ \hline
Policy Neural Network & 4 layer MLP, width 64, ReLU activations \\ \hline
Value Neural Network & 2 layer MLP, width 64, ReLU activations \\ \hline
Optimizer & Rectified Adam, $\beta_1=\beta_2=0.9$ \\ \hline
Learning Rate $\eta$  & $10^{-3}$ \\ \hline
Discount Rate $\gamma$ & $0.99$ \\ \hline
Training Iterations $\numiters[\outer]$ & 10,000 \\ \hline
Batch Size & 32 \\ \hline
Inner interations$\numiters[\inner]$ & 3\\ \hline
\end{tabular}
\label{tab:training}
\end{table}

In order to help speed up learning we also do feature engineering to augment the state space: normalized attacker x coordinate, normalized attacker y coordinate, cos(attacker theta), sin(attacker theta), normalized defender x coordinate, normalized defender y coordinate, cos(defender theta), sin(defender theta), distance between attacker and goal, angular difference between players, distance between players, angular difference between players attacker and goal, angular difference between defender and goal.

\sarjun{}{We approximated \Cref{alg:nested_pgda} and \Cref{alg:sim_pgda} with the REINFORCE algorithm implemented by \Cref{alg:nested_reinforce}, \Cref{alg:sim_reinforce} respectively. Each player is parameterized by a neural network (how to write it?) $\phi_{\inner[][]}$ and $\phi_{\outer[][]}$ which takes as input the augmented state space and outputs a probability distribution over all actions. Each network is four layers of 64 with relu activations.}

\paragraph{Evaluation}
For one of our evaluations, we measure the performance of our trained attacker on a slightly different task to show how the reach-avoid policies can be adapted to obstacle avoidance.

\amy{umm...this is supposed to be the same task that the game was created from. so it should have been described before training. i mean, it should not have been invented just for evaluation.} To demonstrate that our algorithms learn safe policies, we consider the attacker as the ego-agent tasked with reaching a goal in the presence of obstacles. Each obstacle is an instance of a static defender, that is, the defender is fixed in place and does not move.

\sarjun{}{The objective for the attacker is the same as training, reach $\targetstates$ and avoid the $\avoidstates$, however the defender spawns at points depending on the attackers position rather than actively trying to inhibit its progress. In a sense, the Markov game has been converted into a Markov Decision Process  $(\numstates, \states, \targetstates, \safestates, \inneractions,  \initstates, \reward, \detertrans)$} based on the original game without the actions of the defender. Instead, the spawn points of the defender become part of the transition dynamics based on the attacker's state and action and can be though of as static obstacles. 

\amy{i really don't get this. how does this task compare to the training setup?}

We find that the Stackelberg variant outperforms the Nash variant significantly shown in \Cref{tab:eval}. \emph{Reached} corresponds to the attacker reaching $\targetstates$ without ever leaving $\safestates$. A \emph{collision} means the attacker entered $\avoidstates$ and \emph{neither} means that the attacker stayed in $\safestates$ but never reached $\targetstates$.

\begin{center}
\begin{tabular}{|l|l|l|l|}
\hline
 & Reached & Collision & Neither \\
\hline
Stackelberg & 697 & 267 & 36 \\
\hline
Nash & 340 & 606 & 54 \\
\hline
\end{tabular}
\label{tab:eval}
\end{center}


\begin{figure}[htpb]
    \centering
    \begin{subfigure}[t]{0.50\textwidth}
        \includegraphics[width=\textwidth]{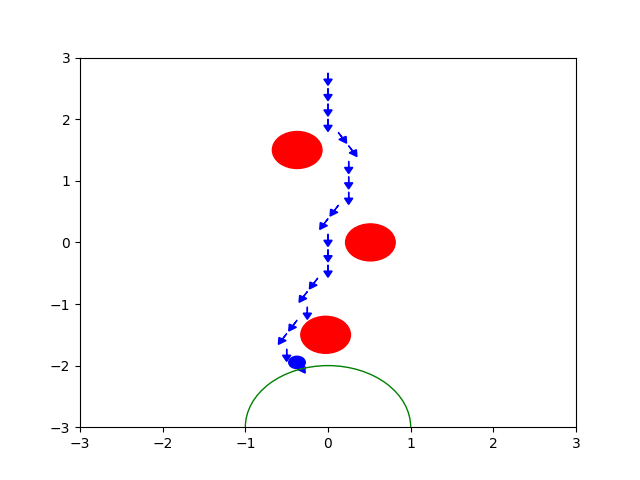}
        \caption{Stackelberg}
    \end{subfigure}
    \hfill 
    \begin{subfigure}[t]{0.49\textwidth}
\includegraphics[width=\textwidth]{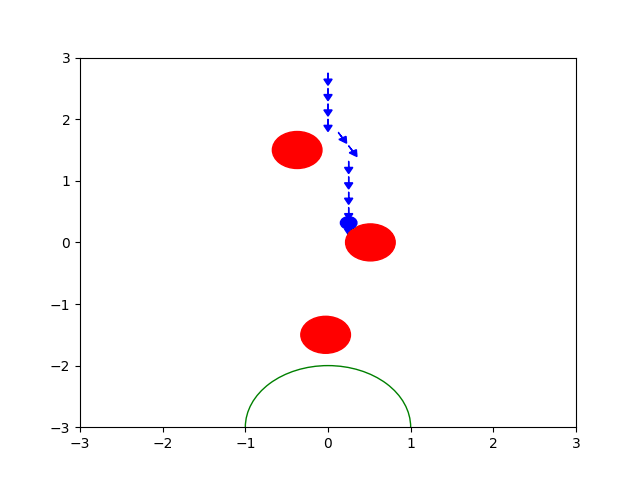}
        \caption{Nash}
    \end{subfigure}
    \caption{Both figures show the an example evaluation run with the same obstacle configuration. The Stackelberg variant successfully reaches the goal while the Nash variant avoids the first obstacle but collides into the second.}
    \label{fig:main}
\end{figure}

\paragraph{Evaluation Setup}
We deploy the evaluation on the same $7 \times 7$ space with the same goal location. The attacker always spawns at $0,3 \text{down}$. Each obstacle is a static defender object with the same capture radius as training. Each obstacle spawns with $x$-coordinate drawn from $U(-1,1)$ and $y$-coordinate $1.5, 0, -1.5$. The are spawned successively according to the attackers $y$-coordinate. As soon as the attacker has successfully navigated by one obstacle, the next one spawns according the the distribution until the episode ends.

\section{Reach-Avoid Games}

\end{document}